\documentclass[thmsa,11pt]{article}%
\usepackage{graphicx}
\usepackage{amssymb}
\usepackage{amsmath}
\usepackage{amsfonts}
\providecommand{\U}[1]{\protect\rule{.1in}{.1in}}
\ifx\pdfoutput\relax\let\pdfoutput=\undefined\fi
\newcount\msipdfoutput
\ifx\pdfoutput\undefined\else
\ifcase\pdfoutput\else
\ifx\paperwidth\undefined\else
\ifdim\paperheight=0pt\relax\else\pdfpageheight\paperheight\fi
\ifdim\paperwidth=0pt\relax\else\pdfpagewidth\paperwidth\fi
\fi\fi\fi

\newtheorem{theorem}{Theorem}[section]

\newtheorem{lemma}[theorem]{Lemma}

\newtheorem{remark}[theorem]{Remark}

\begin{document}

\title{Optimal Learning under Robustness and Time-Consistency\thanks{Department of
Economics, Boston University, lepstein@bu.edu and Zhongtai Securities
Institute of Financial Studies, Shandong University, jsl@sdu.edu.cn. Ji
gratefully acknowledges the financial support of the National Natural Science
Foundation of China (award No. 11571203). We are grateful for suggestions from
two referees and for comments from Tomasz Strzalecki. An earlier version,
titled "Optimal learning and Ellsberg's urns," was posted on arxiv in August
2017.}}
\author{Larry G. Epstein
\and Shaolin Ji}
\maketitle
\date{}

\begin{abstract}
We model learning in a continuous-time Brownian setting where there is prior
ambiguity. The associated model of preference values robustness and is
time-consistent. It is applied to study optimal learning when the choice
between actions can be postponed, at a per-unit-time cost, in order to observe
a signal that provides information about an unknown parameter. The
corresponding optimal stopping problem is solved in closed-form, with a focus
on two specific settings: Ellsberg's two-urn thought experiment expanded to
allow learning before the choice of bets, and a robust version of the
classical problem of sequential testing of two simple hypotheses about the
unknown drift of a Wiener process. In both cases, the link between robustness
and the demand for learning is studied.

\bigskip


\end{abstract}

\newpage

\section{Introduction}

We consider a decision-maker (DM) choosing between three actions whose payoffs
are uncertain because they depend on both exogenous randomness and on an
unknown parameter $\theta$, $\theta=\theta_{0}$ or $\theta_{1}$. She can
postpone the choice of action so as to learn about $\theta$ by observing the
realization of a signal modeled by a Brownian motion with drift. Because of a
per-unit-time cost of sampling, which can be material or cognitive, she faces
an optimal stopping problem. A key feature is that DM does not have sufficient
information to arrive at a single prior about $\theta$, that is, there is
ambiguity about $\theta$. Therefore, prior beliefs are represented by a
nonsingleton set of probability measures, and DM seeks to make robust choices
of both stopping time and action by solving a maxmin problem. In addition, she
is forward-looking and dynamically consistent as in the continuous-time
version of maxmin utility given by Chen and Epstein (2002). One contribution
herein is to extend the latter model to accommodate learning. As a result, we
capture robustness to ambiguity (or model uncertainty), learning and
time-consistency. The other contribution is to investigate optimal learning in
the above setting, with particular focus on two special cases that extend
classical models. The corresponding optimal stopping problems are solved
explicitly and the effects of ambiguity on optimal learning are determined.

The first specific context begins with Ellsberg's metaphorical thought
experiment: \textit{There are two urns, each containing balls that are either
red or blue, where the "known" or risky urn contains an equal number of red
and blue balls, while no information is provided about the proportion of red
balls in the "unknown" or ambiguous urn. DM must choose between betting on the
color drawn from the risky urn or from the ambiguous urn. }The intuitive
behavior highlighted by Ellsberg is the choice to bet on the draw from the
risky urn no matter the color, which behavior is paradoxical for subjective
expected utility theory, or indeed, for any model in which beliefs are
represented by a single probability measure. Ellsberg's paradox is often taken
as a normative critique of the Bayesian model and of the view that the single
prior representation of beliefs is implied by rationality (e.g., Gilboa 2009,
2015; Gilboa et al. 2012). Here we add to the thought experiment by including
a possibility to learn. Specifically, we allow DM to postpone her choice so
that she can observe realizations of a diffusion process whose drift is equal
to the proportion of red in the ambiguous urn. Under specific parametric
restrictions we completely describe the optimal joint learning and betting
strategy. In particular, we show that \textit{it can be optimal to reject
learning completely, and, if some learning is optimal, then it is never
optimal to bet on the risky urn after stopping}. The rationality of no
learning suggests that one needs to reexamine and qualify the common
presumption that ambiguity would fade away, or at least diminish, in the
presence of learning opportunities (Marinacci 2002). It can also explain
experimental findings (Trautman and Zeckhauser 2013) that some subjects
neglect opportunities to learn about an ambiguous urn even at no visible
(material) cost. In addition, our model is suggestive of laboratory
experiments that could provide further evidence on the connection between
ambiguity and the demand for learning.

The second application is to the classical problem of sequential testing of
two simple hypotheses about the unknown drift of a Wiener process. The seminal
papers, both using a discrete-time framework, are Wald (1945,1947), which
shows that the sequential probability ratio test (SPRT) provides an optimal
trade-off between type I and type II errors, and Arrow, Blackwell and Girshick
(1949), which derives SPRT from utility maximization using dynamic programming
arguments. More recently, Peskir and Shiryaev (2006, Ch. 6) employ a Bayesian
subjectivist approach and derive SPRT as the solution to a continuous-time
optimal stopping problem. We extend the latter analysis to accommodate
situations where DM, a statistician/analyst, does not have sufficient
information to justify reliance on a single prior. We show that it is optimal
to stop if every "compatible" Bayesian (one whose prior is an element of the
set of priors used by the robustness-seeking DM) would choose to do so. But
the corresponding statement for "continue" is false: it may be optimal to stop
under robustness even given a realized sample at which all compatible
Bayesians would choose to continue. In this sense, \textit{"sensitivity
analysis" overstates the robustness value of sampling.}

We view our model as normative, which perspective is most evident in the
hypothesis testing context. Time-consistency of preference has obvious
prescriptive appeal. It is important to understand that, roughly speaking,
time-consistency is the requirement that a contingent plan (e.g., a stopping
strategy) that is optimal ex ante remain optimal conditional on every
subsequent realization \emph{assuming there are no surprises or unforeseen
events}. A possible argument against such consistency, (that is sometimes
expressed in the statistics literature), is that surprises are inevitable and
thus that any prescription should take that into account rather than excluding
their possibility. We would agree that a sophisticated decision-maker would
expect that surprises may occur while (necessarily) being unable to describe
what form they could take. However, to the best of our knowledge there
currently does not exist a convincing model in the economics, statistics or
psychology literatures of how such an individual should (or would) behave,
that is, how the awareness that she may be missing something in her perception
of the future should (or would) affect current behavior. That leaves
time-consistency as a sensible guiding principle with the understanding that
reoptimization can (and should) occur if there is a surprise.

A brief review of other relevant literature concludes this introduction. The
classical Bayesian model of sequential decision-making, including in
particular applications to inference and experimentation, are discussed in
Howard (1970) and the references therein. The maxmin model of ambiguity averse
preference is axiomatized in a static setting in Gilboa and Schmeidler (1989),
(which owes an intellectual debt to the Arrow and Hurwicz (1972) model of
decision-making under ignorance), and in a multi-period discrete-time
framework in Epstein and Schneider (2003) where time-consistency is one of the
key axioms. Optimal stopping problems have been studied in the absence of
time-consistency. It is well-known that modeling a concern with ambiguity and
robust decision-making leads to "nonlinear" objective functions, which, in a
dynamic setting and in the absence of commitment, can lead to
time-inconsistency issues (Peskir 2017). A similar issue arises also in a risk
context where there is a known objective probability law, but where preference
does not conform to von Neumann-Morgenstern's expected utility theory (Ebert
and Strack 2018; Huang et al. 2018). Such models are problematic in normative
contexts. It is not clear why one would ever \emph{prescribe} to a
decision-maker (who is unable or unwilling to commit) that she \emph{should}
adopt a criterion function that would imply time-inconsistent plans and that
she \emph{should} then resolve these inconsistencies by behaving strategically
against her future selves (as is commonly assumed). The recursive maxmin model
has been used in macroeconomics and finance (e.g., Epstein and Schneider 2010)
and also in robust multistage stochastic optimization (e.g., Shapiro (2016)
and the references therein, including to the closely related literature on
conditional risk measures). Shapiro focuses on a property of sets of measures,
called rectangularity following Epstein and Schneider (2003), that underlies
recursivity of utility and time-consistency. Most of the existing literature
deals with a discrete-time setting. The theoretical literature on learning
under ambiguity is sparse and limited to passive learning (e.g., Epstein and
Schneider 2007, 2008). With regard to hypothesis testing, this paper adds to
the literature on robust Bayesian statistics (Berger 1984,1985,1994;
Rios-Insua and Ruggeri 2000), which is largely restricted to a static
environment. Walley (1991) goes further and considers both a prior and a
single posterior stage, but not sequential hypothesis testing. For a
frequentist approach to robust sequential testing see Huber (1965).

Closest to the present paper is the literature on bandit problems with
ambiguity and robustness (Caro and Das Gupta 2015; Li 2019). Both papers model
endogenous learning (or experimentation) by maxmin dynamically consistent
agents. Their models differ from ours in that they assume discrete time, an
exogenously given horizon, and also in the nature of experimentation. In our
model, the once-and-for-all choice of action and resulting payoff come after
all learning has ceased, while in bandit problems, action choice and flow
payoffs are continuous and intertwined with learning (for example, the cost of
experimentation is the implied reduction in current flow payoffs).
Consequently, their analyses and characterizations are much different, for
example, their focus on the existence of a suitable Gittins index has no
counterpart in our model.

The paper proceeds as follows. The next section describes the model of utility
extending Chen-Epstein to accommodate learning. Readers who are primarily
interested in applications can skip this relatively technical section and move
directly to \S 3 where the "applied" optimal stopping problems are studied.
The (more) general optimal stopping problem is solved in \S 4 (Theorem
\ref{thm-general}), thereby providing a unifying perspective on the two
applications and some indication of the robustness of the results therein.
Proofs are contained in the e-companion to this paper.

\section{Recursive utility with learning\label{section-rcu}}

For background regarding time-consistency in the maxmin framework, consider
first the following informal outline that anticipates the specific setting of
this paper. DM faces uncertainty about a payoff-relevant state space $\Omega$
due to uncertainty about the value of a parameter $\theta\in\Theta$. Each
$\theta$ determines a unique probability law on $\Omega$, but there is prior
ambiguity about the parameter that is represented by a nonsingleton set
$\mathcal{M}_{0}$ of priors on $\Theta$. As time proceeds, DM learns about the
parameter through observation of a signal whose increments are distributed
i.i.d. conditional on $\theta$. At issue is how to model beliefs about
$\Omega$, that is, the set $\mathcal{P}_{0}$ of predictive priors. (Throughout
we adopt the common practice of distinguishing terminologically between
beliefs about the state space, referred to as predictive priors, and beliefs
about parameters, which are referred to as priors.) A seemingly natural
approach is to take $\mathcal{P}_{0}$ to be the set of all measures that can
be obtained by combining some prior $\mu_{0}$ in $\mathcal{M}_{0}$ with the
given conditionally i.i.d. likelihood. Learning is modeled through the set of
posteriors $\mathcal{M}_{t}$ at $t$ obtained via prior-by-prior Bayesian
updating of $\mathcal{M}_{0}$, and a corresponding set $\mathcal{P}_{t}$ of
predictive posteriors is obtained as above. Finally, at each $t\geq0$,
$\mathcal{P}_{t}$ guides choice according to the maxmin model. The point,
however, is that time-consistency is violated: in general, ex ante optimal
plans do not remain optimal according to updated beliefs. The reason is
straightforward. Behavior at $t$\ is depends on the worst-case posterior
$\mu_{t}$\ in $\mathcal{M}_{t}$, but worst-cases at different nodes need not
belong to same prior $\mu_{0}$. This is in contrast with the ex ante
perspective expressed via{\small \ }$\mathcal{P}_{0}${\small \ }where a single
worst-case{\small \ }prior $\mu_{0}$ determines the entire ex ante optimal
plan. To restore dynamic consistency, one can enlarge $\mathcal{P}_{0}$\ by
adding to it all measures obtained by pasting together alien posteriors,
leading to a "rectangular" set that is closed with respect to further pasting.
One can think of the enlarged set as capturing \emph{both}\ the subjectively
possible probability laws \emph{and} backward induction reasoning by DM.a

See Epstein and Schneider (2003) for further discussion and axiomatic
foundations in a discrete-time framework, and Chen and Epstein (2002)--CE
below--for a continuous-time formulation that we outline next. Then we
describe how it can be adapted to include learning with partial information.
The latter description is given in the simplest context adequate for the
applications below. However, it should be clear that it can be adapted more generally.

Let $(\Omega,\mathcal{G}_{\infty},P_{0})$ be a probability space, and
$W=(W_{t})_{0\leq t<\infty}$ a $1$-dimensional Brownian motion which generates
the filtration $\mathcal{G}=\{\mathcal{G}_{t}\}_{t\geq0}$, with $\mathcal{G}%
_{t}\nearrow\mathcal{G}_{\infty}$. (All probability spaces are taken to be
complete and all related filtrations are augmented in the usual sense.) The
measure $P_{0}$ is a reference measure whose role is only to define null
events. CE define a set of predictive priors $\mathcal{P}_{0}$ on
$(\Omega,\mathcal{G}_{\infty})$ through specification of their densities with
respect to $P_{0}$. To do so, they take as an additional primitive a (suitably
adapted) set-valued process $\left(  \Xi_{t}\right)  $. (Technical
restrictions are that $\Xi_{t}:\Omega\rightsquigarrow K\subset\mathbb{R}^{d}$
for some compact set $K$ independent of $t$, $0\in\Xi_{t}\left(
\omega\right)  $ $dt\otimes dP_{0}\ a.s.$, and that each $\Xi_{t} $ is convex-
and compact-valued.) Define the associated set of real-valued processes by
\[
\Xi=\{\eta=(\eta_{t})\mid\eta_{t}(\omega)\in\Xi_{t}(\omega)\;dt\otimes
dP_{0}\ a.s.\}.
\]
Then each $\eta\in\Xi$ \ defines a probability measure on $\mathcal{G}%
_{\infty}$, denoted $P^{\eta}$, that is equivalent to $P_{0}$ on each
$\mathcal{G}_{t}$, and is given by%
\[
\frac{dP^{\eta}}{dP_{0}}\mid_{\mathcal{G}_{t}}=\exp\{-\int\nolimits_{0}%
^{t}\eta_{s}^{2}ds-\int\nolimits_{0}^{t}\eta_{s}dW_{s}\}\text{ for all
}t\text{.}%
\]
Accordingly, each $\eta_{t}(\omega)\in\Xi_{t}(\omega)$ can be thought of
roughly as defining conditional beliefs about $\mathcal{G}_{t+dt}$, and
$\Xi_{t}\left(  \omega\right)  $ is called the set of \emph{density
generators} at $\left(  t,\omega\right)  $. By the Girsanov Theorem,
\begin{equation}
dW_{t}^{\eta}=\eta_{t}dt+dW_{t}\label{W'}%
\end{equation}
is a Brownian motion under $P^{\eta}$, which thus can be understood as an
alternative hypothesis about the drift of the driving process $W$ (the drift
is $0$ under $P_{0}$). Finally,
\begin{equation}
\mathcal{P}_{0}\equiv\left\{  P^{\eta}:\eta\in\Xi\right\}  \text{.}%
\label{Pcal0}%
\end{equation}
(The "pasting" referred to above is accomplished through the fact that $\Xi$
is constructed by taking all selections from the $\Xi_{t}$s.)

The set $\mathcal{P}_{0}$ is used to define a time $0$ utility function on a
suitable set of random payoffs denominated in utils. In order to model in the
sequel the choice of how long to learn (or sample), we consider a set of
stopping times $\tau$, that is, each $\tau$ is an adapted $\mathbb{R}_{+}%
$-valued and $\{\mathcal{G}_{t}\}$-adapted random variable defined on $\Omega
$, that is, $\{\omega:\tau\left(  \omega\right)  >t\}\in$ $\mathcal{G}_{t}$
for every $t$. For each such $\tau$, utility is defined on the set $L(\tau) $
of real-valued random variables given by%
\[
L(\tau)=\{\xi\mid\xi\text{ is }\mathcal{G}_{\tau}\text{-measurable and
}\underset{Q\in\mathcal{P}_{0}}{\sup}E_{Q}\mid\xi\mid<\infty\}\text{.}%
\]

The time $0$ utility of any $\xi\in L(\tau)$ is given by%
\begin{equation}
U_{0}\left(  \xi\right)  =\underset{Q\in\mathcal{P}_{0}}{\inf}E_{Q}%
\xi=-\underset{Q\in\mathcal{P}_{0}}{\sup}E_{Q}[-\xi].\label{U0}%
\end{equation}
It is natural to consider also conditional utilities at each $\left(
t,\omega\right)  $, where
\begin{equation}
U_{t}\left(  \xi\right)  =\underset{Q\in\mathcal{P}_{0}}{\text{ess}\inf}%
E_{Q}[\xi\mid\mathcal{G}_{t}]\text{.}\label{Ut}%
\end{equation}
In words, $U_{t}\left(  \xi\right)  $ is the utility of $\xi$ at time $t$
conditional on the information available then\ and given the state $\omega$
(the dependence of $U_{t}\left(  \xi\right)  $ on $\omega$ is suppressed
notationally). The special construction of $\mathcal{P}_{0}$ delivers the
following counterpart of the law of total probability (or law of iterated
expectations): For each $\xi$, and $0\leq t<t^{\prime}$,%
\begin{equation}
U_{t}\left(  \xi\right)  =\underset{Q\in\mathcal{P}_{0}}{\text{ess}\inf}%
E_{Q}\left[  U_{t^{\prime}}\left(  \xi\right)  \mid\mathcal{G}_{t}\right]
\text{.}\label{LIE}%
\end{equation}
This recursivity ultimately delivers the time-consistency of optimal choices.

The components $P_{0}$, $W$, $\left(  \Xi_{t}\right)  $ and $\{\mathcal{G}%
_{t}\}$ are primitives in CE. Next we specify them in terms of the deeper
primitives of a model that includes learning about an unknown parameter
$\theta\in\Theta\subset\mathbb{R}$.

Specifically, begin with a measurable space $\left(  \Omega,\mathcal{F}%
\right)  $, a filtration $\{\mathcal{F}_{t}\}$, $\mathcal{F}_{t}%
\nearrow\mathcal{F}_{\infty}\subset\mathcal{F}$, and a collection $\{P^{\mu
}:\mu\in\mathcal{M}_{0}\}$ of pairwise equivalent probability measures on
$\left(  \Omega,\mathcal{F}\right)  $. Though $\theta$\ is an unknown
deterministic parameter, for mathematical precision we view $\theta$\ as a
random variable on $\left(  \Omega,\mathcal{F}\right)  $. Further, for each
$\mu\in\mathcal{M}_{0}$, $P^{\mu}$ induces the distribution $\mu$ for $\theta$
via $\mu(A)=P^{\mu}(\{\theta\in A\})$\ for all Borel measurable $A\subset
\Theta$. Accordingly, $\mathcal{M}_{0}$ can be viewed as a set of priors on
$\Theta$, and its nonsingleton nature indicates ambiguity about $\theta$.
There is also a standard Brownian motion $B=(B_{t})$, with generated
filtration $\{\mathcal{F}_{t}^{B}\}$, such that $B$ is independent of $\theta$
under each $P^{\mu}$. $B$ is the Brownian motion driving the signals process
$Z=(Z_{t})$ according to%
\begin{equation}
Z_{t}=\int\nolimits_{0}^{t}\theta ds+\int\nolimits_{0}^{t}\sigma dB_{s}=\theta
t+\sigma B_{t},\label{Z}%
\end{equation}
where $\sigma$ is a known positive constant. Because only realizations of
$Z_{t}$ are observable, take $\{\mathcal{G}_{t}\}$ to be the filtration
generated by $Z$. Assuming knowledge of the signal structure, Bayesian
updating of $\mu\in\mathcal{M}_{0}$ gives the posterior $\mu_{t}$ at time $t$.
Thus prior-by-prior Bayesian updating leads to the set-valued process
$(\mathcal{M}_{t})$ of posteriors on $\theta$.

Proceed to specify the other CE components $P_{0}$, $W$ and $\left(  \Xi
_{t}\right)  $.

\medskip

\noindent\textbf{Step 1.} Take $\mu\in\mathcal{M}_{0}$. By standard filtering
theory (Liptser and Shiryaev 1977, Theorem 8.3), if we replace the unknown
parameter $\theta$ by the estimate $\widehat{\theta}_{t}^{\mu}=\int\theta
d\mu_{t}$, then we can rewrite (\ref{Z}) in the form%
\begin{align}
dZ_{t}  & =\hat{\theta}_{t}^{\mu}\left(  Z_{t}\right)  dt+\sigma(dB_{t}%
+\frac{\theta-\hat{\theta}_{t}^{\mu}\left(  Z_{t}\right)  }{\sigma
}dt)\label{Z2-0}\\
& =\hat{\theta}_{t}^{\mu}\left(  Z_{t}\right)  dt+\sigma d\tilde{B}_{t}^{\mu
}\text{,}\nonumber
\end{align}
where the innovation process $(\tilde{B}_{t}^{\mu})$ is a standard
$\{\mathcal{G}_{t}\}$-adapted Brownian motion on $(\Omega,\mathcal{G}_{\infty
},P^{\mu})$. Thus $(\tilde{B}_{t}^{\mu})$ takes the same role as $(W_{t}%
^{\eta})$ in CE (see (\ref{W'}) above). Rewrite (\ref{Z2-0}) as
\[
d\tilde{B}_{t}^{\mu}=-\frac{1}{\sigma}\hat{\theta}_{t}^{\mu}\left(
Z_{t}\right)  dt+\frac{1}{\sigma}dZ_{t}%
\]
which suggests that $\left(  Z_{t}/\sigma\right)  $ (resp. $(-\hat{\theta}%
_{t}^{\mu}\left(  Z_{t}\right)  /\sigma)$) can be chosen as the Brownian
motion $(W_{t})$ (resp. the drift $(\eta_{t})$) in (\ref{W'}).

\medskip

\noindent\textbf{Step 2.} Find a reference probability measure $P_{0}$ on
$(\Omega,\mathcal{G}_{\infty})$ under which $\left(  Z_{t}/\sigma\right)  $ is
a $\{\mathcal{G}_{t}\}$-adapted Brownian motion on $(\Omega,\mathcal{G}%
_{\infty})$. Fix $\overline{\mu}\in\mathcal{M}_{0}$ and define $P_{0}$ by:
\[%
\begin{array}
[c]{rl}%
\frac{dP_{0}}{dP^{\overline{\mu}}}\mid_{\mathcal{G}_{t}} & =\exp\{-\frac
{1}{2\sigma^{2}}\int\nolimits_{0}^{t}(\hat{\theta}_{s}^{\overline{\mu}}\left(
Z_{s}\right)  )^{2}ds-\frac{1}{\sigma}\int\nolimits_{0}^{t}\hat{\theta}%
_{s}^{\overline{\mu}}\left(  Z_{s}\right)  d\tilde{B}_{s}^{\overline{\mu}}\}\\
& =\exp\{\frac{1}{2\sigma^{2}}\int\nolimits_{0}^{t}(\hat{\theta}%
_{s}^{\overline{\mu}}\left(  Z_{s}\right)  )^{2}ds-\frac{1}{\sigma^{2}}%
\int\nolimits_{0}^{t}\hat{\theta}_{s}^{\overline{\mu}}\left(  Z_{s}\right)
dZ_{s}\}\text{.}%
\end{array}
\]
By Girsanov's Theorem, $\left(  Z_{t}/\sigma\right)  $ is a $\{\mathcal{G}%
_{t}\}$-adapted Brownian motion under $P_{0}$.

\medskip\medskip

\noindent\textbf{Step 3.} Viewing $P_{0}$ as a reference measure, perturb it.
For each $\mu\in\mathcal{M}_{0}$, define $P_{0}^{\mu}$ on $(\Omega
,\mathcal{G}_{\infty})$ by%
\[
\frac{dP_{0}^{\mu}}{dP_{0}}\mid_{\mathcal{G}_{t}}=\exp\{-\frac{1}{2\sigma^{2}%
}\int\nolimits_{0}^{t}(\hat{\theta}_{s}^{\mu}\left(  Z_{s}\right)
)^{2}ds+\frac{1}{\sigma^{2}}\int\nolimits_{0}^{t}\hat{\theta}_{s}^{\mu}\left(
Z_{s}\right)  dZ_{s}\}.
\]
By Girsanov, $d\tilde{B}_{t}^{\mu}=-\frac{1}{\sigma}\hat{\theta}_{t}^{\mu
}\left(  Z_{t}\right)  dt+\frac{1}{\sigma}dZ_{t}$ is a Brownian motion under
$P_{0}^{\mu}$.

In general, $P^{\mu}\not =P_{0}^{\mu}$. However, they induce the identical
distribution for $Z$. This is because $(\tilde{B}_{t}^{\mu})$ is a
$\{\mathcal{G}_{t}\}$-adapted Brownian motion under both $P^{\mu}$ and
$P_{0}^{\mu}$. Therefore, by the uniqueness of weak solutions to SDEs, the
solution $Z_{t}$ of (\ref{Z2-0}) on $(\Omega,\mathcal{F}_{\infty},P^{\mu}) $
and the solution $Z^{\prime}$ of (\ref{Z2-0}) on $(\Omega,\mathcal{G}_{\infty
},P_{0}^{\mu})$ have identical distributions. (Argue as in Oksendal (2005,
Example 8.6.9). Given that only the distribution of signals matters in our
model, there is no reason to distinguish between the two probability measures.
Thus we apply CE to the following components: $W$ and $P_{0}$ defined in Step
2, and $\Xi_{t}$ given by
\begin{equation}
\Xi_{t}=\{-\hat{\theta}_{t}^{\mu}/\sigma:\mu\in\mathcal{M}_{0},\widehat{\theta
}_{t}^{\mu}=\int\theta d\mu_{t}\}\text{.}\ \label{THETAHAT}%
\end{equation}

In summary, taking these specifications for $P_{0}$, $W$, $\left(  \Xi
_{t}\right)  $ and $\{\mathcal{G}_{t}\}$ in the CE model yields a set
$\mathcal{P}_{0}$ of predictive priors, and a corresponding utility function,
that capture prior ambiguity about the parameter $\theta$ (through
$\mathcal{M}_{0}$), learning as signals are realized (through updating to the
set of posteriors $\mathcal{M}_{t}$), and robust (maxmin) and time-consistent
decision-making (because of (\ref{LIE})). We use this model in the optimal
stopping problems that follow. The only remaining primitive is $\mathcal{M}%
_{0}$, which is specified to suit the particular setting of interest.

As indicated, the key technical step in our extension of CE is in adopting the
weak formulation rather than their strong formulation. For readers who may be
unfamiliar with this distinction we suggest Oksendal (2005, Section 5.3) for
discussion of weak versus strong solutions of SDEs, and Zhang (2017, Chapter
9). The latter exposits both the technical advantages of the weak formulation
and its economic rationale, notably in models with imperfect information (such
as here, where given (\ref{Z}), $Z$ is observed but not $B$), or asymmetric
information (such as in principal-agent models). In our context, the weak
formulation is suggested if one views $B$ not as modeling a physical noise or
shock, but rather as a way to specify that the \emph{distribution} of $\left(
Z_{t}-\theta t\right)  /\sigma$ is standard normal (conditional on $\theta$).

\section{Optimal learning}

\subsection{The framework and general problem\label{section-general}}

DM must choose an action from the set $A=\{a_{0},a_{1},a_{2}\}$. Payoffs are
uncertain and depend on an unknown parameter $\theta$. Before choosing an
action, DM can learn about $\theta$ by observing realizations of the signal
process $Z$ given by (\ref{Z}), where $\sigma$ is a known positive constant.
There is a constant per-unit-time cost $c>0$ of learning. (The underlying
state space $\Omega$, the filtration $\{\mathcal{G}_{t}\}$ generated by $Z$,
and other notation are as in \S 2. Unless specified otherwise, all processes
below are taken to be $\{\mathcal{G}_{t}\}$-adapted even where not stated explicitly.)

If DM stops learning at $t$, then her conditional expected payoff (in utils)
is $X_{t}$; think of $X_{t}$ as the indirect utility she can attain by
choosing optimally from $A$. DM is forward-looking and has time $0$ beliefs
about future signals given by the set $\mathcal{P}_{0}\subset\Delta\left(
\Omega,\mathcal{G}_{\infty}\right)  $ described in the previous section. Her
choice of when to stop is described by a stopping time (or strategy) $\tau$,
which is restricted to be uniformly integrable ($\sup_{Q\in\mathcal{P}_{0}%
}E_{Q}\tau<\infty$); the set of all stopping strategies is $\Gamma$. As a
maxmin agent she chooses an optimal stopping strategy $\tau^{\ast}$ by solving%
\begin{equation}
\max_{\tau\in\Gamma}\min_{P\in\mathcal{P}_{0}}E_{P}\left(  X_{\tau}%
-c\tau\right)  \text{.}\label{stopX}%
\end{equation}
It remains to specify $\mathcal{M}_{0}$, which determines $\mathcal{P}_{0}$ as
described in \S 2, and $X_{t}$.

We assume that all priors $\mu$ in $\mathcal{M}_{0}$ have binary support
$\Theta=\{\theta_{0}$,$\theta_{1}\}$, $\theta_{0}<\theta_{1}$. Specifically,
let%
\begin{equation}
\mathcal{M}_{0}=\{\mu^{m}=(1-m)\delta_{\theta_{0}}+m\delta_{\theta_{1}%
}:\underline{m}_{0}\leq m\leq\overline{m}_{0}\}\text{.}\label{M0}%
\end{equation}
Therefore, $\mathcal{M}_{0}$ can be identified with the probability interval
$\left[  \underline{m}_{0},\overline{m}_{0}\right]  $ for the larger parameter
value $\theta_{1}$. Let $0<\underline{m}_{0}<\overline{m}_{0}<1$.

Bayesian updating of each prior yields the following set of posteriors at $t$,%
\begin{equation}
\mathcal{M}_{t}=\{(1-m)\delta_{\theta_{0}}+m\delta_{\theta_{1}}:\underline{m}%
_{t}\leq m\leq\overline{m}_{t}\}\text{,}\label{Mt}%
\end{equation}
where, by Liptser and Shiryaev (1977, Theorem 9.1),
\begin{equation}
\underline{m}_{t}=\frac{\frac{\underline{m}_{0}}{1-\underline{m}_{0}}%
\varphi(t,Z_{t})}{1+\frac{\underline{m}_{0}}{1-\underline{m}_{0}}%
\varphi(t,Z_{t})}\text{, }\overline{m}_{t}=\frac{\frac{\overline{m}_{0}%
}{1-\overline{m}_{0}}\varphi(t,Z_{t})}{1+\frac{\overline{m}_{0}}%
{1-\overline{m}_{0}}\varphi(t,Z_{t})}\text{,}\label{mt}%
\end{equation}
and%
\begin{equation}
\varphi(t,z)=\exp\{\frac{\theta_{1}-\theta_{0}}{\sigma^{2}}z-\frac{1}%
{2\sigma^{2}}(\theta_{1}^{2}-\theta_{0}^{2})t\}.\label{phi}%
\end{equation}

Conditional on the parameter value, payoffs are given by $u\left(
a_{i},\theta_{j}\right)  $, where each $u\left(  a_{i},\theta_{j}\right)  $ is
nonnegative. Think of $u\left(  \cdot,\theta_{j}\right)  $ as including the
valuation of any risk remaining even if $\theta_{j}$ is known to be true, for
example, $u\left(  a_{i},\theta_{j}\right)  $ could be the expected utility of
the lottery implied by $\left(  a_{i},\theta_{j}\right)  $. Payoffs are
assumed to satisfy: for each $i,j=0,1$, $i\not =j$,
\begin{equation}
u\left(  a_{j},\theta_{j}\right)  =u\left(  a_{i},\theta_{i}\right)  >u\left(
a_{j},\theta_{i}\right)  \text{.}\label{u}%
\end{equation}
Thus $a_{0}$ is better than $a_{1}$ given $\theta_{0}$, and the reverse given
$\theta_{1}$, and the payoff to the better action is the same for both
parameter values. The payoff to the third action $a_{2}$ does not depend on
$\theta$, and can be thought of as a default or outside option. Its payoff is
not ambiguous because incomplete confidence about $\theta$ is the only source
of ambiguity in the model, but choice of $a_{2}$ may entail risk. Adopt the
notation
\begin{equation}
u_{2}=u\left(  a_{2},\theta_{0}\right)  =u\left(  a_{2},\theta_{1}\right)
\text{.}\label{u0}%
\end{equation}
It is evident that action $a_{2}$ may be irrelevant if its payoff is
sufficiently low, for example, if $u_{2}=0$. To exclude the trivial case where
$a_{2}$ is always chosen, assume that%
\[
u_{2}<u\left(  a_{i},\theta_{i}\right)  \text{, \ }i=0,1\text{.}%
\]

Consider next payoffs conditional on time $t$ beliefs about $\theta$ as
represented by the set of posteriors $\mathcal{M}_{t}$. The Gilboa-Schmeidler
utility of $a_{i}$ is $\min_{\mu\in\mathcal{M}_{t}}\int u\left(  a_{i}%
,\theta\right)  d\mu$. Therefore, if DM chooses an optimal action at time $t$,
then her payoff is
\begin{equation}
X_{t}=\max\left\{  \min_{\mu\in\mathcal{M}_{t}}\int u\left(  a_{0}%
,\theta\right)  d\mu,\min_{\mu\in\mathcal{M}_{t}}\int u\left(  a_{1}%
,\theta\right)  d\mu,u_{2}\right\}  \text{.}\label{Xt}%
\end{equation}

The preceding completes specification of the optimal stopping problem
(\ref{stopX}). Its solution is described in \S 4 under two \emph{alternative}
additional assumptions:

\begin{description}
\item[Payoff symmetry] $u\left(  a_{0},\theta_{1}\right)  =u\left(
a_{1},\theta_{0}\right)  $

\item[No risky option] $u_{2}\leq u\left(  a_{i},\theta_{j}\right)  $,
$i\not =j=0,1$
\end{description}

\noindent The first assumption adds to the symmetry contained in (\ref{u}).
Given (\ref{u}), the second implies that action $a_{2}$ is (weakly) inferior
to each of $a_{0}$ and $a_{1}$ conditional on either parameter value. Hence,
it would never be chosen uniquely and can be ignored, leaving only two
actions. These assumptions are satisfied respectively by the two special
models upon which we focus: Ellsberg's urns (payoff symmetry) and hypothesis
testing (no risky option). We focus on these first because they extend classic
models in the literature and because they provide simply distinct insights
into the connection between ambiguity and optimal learning.

\subsection{Learning and Ellsberg's urns\label{section-ellsberg}}

There are two urns each containing balls that are either red or blue: a risky
urn in which the proportion of red balls is $\frac{1}{2}$ and an ambiguous urn
in which the color composition is unknown. Denote by $\theta+\frac{1}{2}$ the
unknown proportion of red balls. Thus $\theta$ denotes the bias towards red:
$\theta>0$ indicates more red than blue, $\theta<0$ indicates the opposite,
and $\theta=0$ indicates an equal number as in the risky urn. DM can choose
between betting on the draw from the risky or ambiguous urn and also on
drawing red or blue. In the absence of learning, the intuitive behavior
highlighted by Ellsberg is to bet on the draw from the risky urn no matter the
color. Here we consider betting preference when an ambiguity averse
decision-maker can defer the choice between bets until after learning
optimally about $\theta$.

To do so, we apply the model described above with particular specifications
for its key primitives $A$, $\Theta$, $\mathcal{M}_{0}$ and $u$. For $A$, let
$a_{2}$ denote a bet on the risky urn and let $a_{1}$ ($a_{0}$) denote the bet
on drawing red (blue) from the ambiguous urn. (Note that there is no need to
differentiate between bets on red and blue for the risky urn.) Take
$\Theta=\{\theta_{0},\theta_{1}\}$, where $\theta_{0}+\theta_{1}=0$, or
equivalently, for some $0<\alpha<\frac{1}{2}$,%
\begin{equation}
\theta_{0}=-\alpha\text{, }\theta_{1}=\alpha\text{. }\label{alpha}%
\end{equation}
Thus only two possible biases, of equal size, are thought possible, (the
proportion of red is either $\frac{1}{2}-\alpha$ or $\frac{1}{2}+\alpha$).
However, there is ambiguity about which direction for the bias is more likely.
This ambiguity is modeled by $\mathcal{M}_{0}$ having the form in (\ref{M0}),
where we assume in addition that the probability interval for $\alpha$ (the
bias towards red) is such that $\underline{m}_{0}+\overline{m}_{0}=1$, or
equivalently, for some $0<\epsilon<1$,
\begin{equation}
\underline{m}_{0}=\frac{1-\epsilon}{2}\text{, }\overline{m}_{0}=\frac
{1+\epsilon}{2}\text{.}\label{epsilon}%
\end{equation}
Thus the lowest probability for a bias towards blue equals that for red,
implying indifference at time $0$ between bets on red and blue. This
assumption, and also the color symmetry in (\ref{alpha}), are natural since
information about the ambiguous urn gives no reason to distinguish between colors.

We are left with the two parameters $\alpha$ and $\epsilon$. We interpret
$\epsilon$ as modeling ambiguity (aversion): the probability interval $\left[
\frac{1-\epsilon}{2},\frac{1+\epsilon}{2}\right]  $ for the bias towards red
is larger if $\epsilon$ increases. At the extreme when $\epsilon=0$, then
$\mathcal{M}_{0}$ is the singleton according to which the two biases are
equally likely, and DM is a Bayesian who faces uncertainty with variance
$\alpha^{2}$ about the true bias,$\,\ $but no ambiguity. We interpret $\alpha$
as measuring the degree of this prior uncertainty, or \emph{prior variance};
($\alpha=0$ implies certainty that the composition of the ambiguous urn is
identical to that of the risky urn).

Finally, specify payoffs $u$. All bets have the same winning and losing
prizes, denominated in utils, which can be normalized to $1$ and $0$
respectively. Given the composition of the ambiguous urn, then only risk is
involved in every bet, and an expected utility calculation yields%
\begin{equation}
u\left(  a_{0},-\alpha\right)  =u\left(  a_{1},\alpha\right)  =\alpha
+\tfrac{1}{2}\text{, }u\left(  a_{0},\alpha\right)  =u\left(  a_{1}%
,-\alpha\right)  =\alpha-\tfrac{1}{2}\text{, and }u_{2}=\tfrac{1}{2}%
\text{.}\label{u-ellsberg}%
\end{equation}
The assumptions in \S 3.1 are readily verified.

For convenience of the reader, we include the implied expression for the
conditional payoff $X_{t}=X(Z_{t})$:%

\begin{equation}
X(Z_{t})=\left\{
\begin{array}
[c]{cc}%
(\tfrac{1}{2}+\alpha)-\frac{2\alpha}{1+\frac{1-\epsilon}{1+\epsilon}%
\varphi(Z_{t})} & \text{if }Z_{t}>\frac{\sigma^{2}}{2\alpha}\log
(\frac{1+\epsilon}{1-\epsilon})\\
(\tfrac{1}{2}-\alpha)+\frac{2\alpha}{1+\frac{1+\epsilon}{1-\epsilon}%
\varphi(Z_{t})} & \text{if }Z_{t}<-\frac{\sigma^{2}}{2\alpha}\log
(\frac{1+\epsilon}{1-\epsilon})\\
\frac{1}{2} & \text{otherwise,}%
\end{array}
\right. \label{Xt-ellsberg}%
\end{equation}
where $\varphi(z)=\exp\left(  2\alpha z/\sigma^{2}\right)  $. Thus if $Z_{t} $
is large positive (negative), then a bet on drawing red (blue) from the
ambiguous urn is optimal. For intermediate values, there is not enough
evidence for a bias in either direction to compensate for the ambiguity and
betting on the risky urn is optimal. This is true in particular ex ante where
$Z_{0}=0$, consistent with the intuitive ambiguity-averse behavior in
Ellsberg's 2-urn experiment without learning.

We give an explicit solution to the optimal stopping problem (\ref{stopX})
satisfying (\ref{alpha})-(\ref{u-ellsberg}). To do so, let%
\begin{equation}
l(r)=2\log(\frac{r}{1-r})-\frac{1}{r}+\frac{1}{1-r},\text{\ }r\in
(0,1)\text{,}\label{el}%
\end{equation}
and define $\widehat{r}$ by%
\begin{equation}
l(\widehat{r})=\frac{2\alpha^{3}}{c\sigma^{2}}\text{.}\label{rhat}%
\end{equation}
$\widehat{r}$ is uniquely defined thereby and $\frac{1}{2}<\widehat{r}<1$,
because $l(\cdot)$ is strictly increasing, $l(0)=-\infty$, $l(\frac{1}{2})=0$,
and $l(1)=\infty$. \

\begin{theorem}
\label{thm-ellsberg}(i) $\tau^{\ast}=0$ if and only if $\frac{1+\epsilon}%
{2}\geq\widehat{r}$, in which case $X_{\tau^{\ast}}=X_{0}=\frac{1}{2}$.

\noindent(ii) Let $\frac{1+\epsilon}{2}<\widehat{r}$. Then the optimal
stopping time satisfies $\tau^{\ast}>0$ and is given by%
\[
\tau^{\ast}=\min\{t\geq0:\text{ }\mid Z_{t}\mid\geq\overline{z}\},
\]
where
\begin{equation}
\overline{z}=\frac{\sigma^{2}}{2\alpha}\left[  \log\frac{1+\epsilon
}{1-\epsilon}+\log\frac{\overline{r}}{1-\overline{r}}\right]  >0\text{,}%
\label{zbar}%
\end{equation}
and $\overline{r}$, $\widehat{r}<\overline{r}<1$, is the unique solution to
the equation
\begin{equation}
l(r)+l(\frac{1+\epsilon}{2})=\frac{4\alpha^{3}}{c\sigma^{2}}\text{.}%
\label{rbar}%
\end{equation}
Moreover, on stopping either the bet on red is chosen (if $Z_{\tau^{\ast}}%
\geq\overline{z}$) or the bet on blue is chosen (if $Z_{\tau^{\ast}}%
\leq-\overline{z}$); the bet on the risky urn is never optimal at $\tau^{\ast
}>0$. Finally, if $\epsilon<\epsilon^{\prime}<2\widehat{r}-1$, and if
$\tau^{\ast\prime}$ is the corresponding optimal stopping time, then
$\tau^{\ast\prime}\geq\tau^{\ast}$.
\end{theorem}

The two cases are defined by the relative magnitudes of $\epsilon$,
parametrizing ambiguity, and $\widehat{r}$, which is an increasing function of
$\alpha^{3}/\left(  c\sigma^{2}\right)  $; in particular, through $\alpha$, it
depends positively on the payoff to knowing the direction of the true bias.
Thus (i) considers the case where ambiguity is large realtive to payoffs (and
taking also sampling cost and signal variance into account). Then no learning
is optimal and the bet on the risky urn is chosen immediately. In contrast,
some learning is necessarily optimal given small ambiguity (case (ii)),
including in the limiting Bayesian model with $\epsilon=0$. Thus \textit{it is
optimal to reject learning if and only if ambiguity, as measured by }%
$\epsilon$\textit{, is suitably large}. In case (ii), it is optimal to sample
as long as the signal $Z_{t}$ lies in the continuation interval $\left(
-\overline{z},\overline{z}\right)  $. Two features of this learning region
stand out. First, when $Z_{t}$ hits either endpoint, learning stops and DM
bets on the ambiguous urn. Thus the \textit{risky urn is chosen (if and) only
if it is not optimal to learn}. The second noteworthy feature is that
\textit{sampling increases with greater ambiguity} as measured by $\epsilon$,
though when $\epsilon$ reaches $2\widehat{r}-1$, then, by (i), it is optimal
to reject any learning.

There is simple intuition for the preceding. First, consider the effect of
ambiguity (large $\epsilon$) on the incentive to learn. DM's prior beliefs
admit only $\alpha$ and $-\alpha$ as the two possible values for the true
bias. She will incur the cost of learning if she believes that she is likely
to learn quickly which of these is true. She understands that she will come to
accept $\alpha$ (or $-\alpha$) as being true given realization of sufficiently
large positive (negative) values for $Z_{t}$. A difficulty is that she is not
sure which probability law in her set $\mathcal{P}_{0}$ describes the signal
process. As a conservative decision-maker, she bases her decisions on the
worst-case scenario $P^{\ast}$ in her set. Because she is trying to learn, the
worst-case minimizes the probability of extreme, hence revealing, signal
realizations, which, informally speaking, occurs if $P^{\ast}(\{dZ_{t}>0\}\mid
Z_{t}>0)$\ and $P^{\ast}(\{dZ_{t}<0\}\mid Z_{t}<0)${\small \ }are as small as
possible. That is, if $Z_{t}>0$, then the distribution of the increment
$dZ_{t}$ is computed using the posterior associated with that prior in
$\mathcal{M}_{0}$ which assigns the largest probability $\frac{1+\epsilon}{2}$
to the negative bias $-\alpha$, while if $Z_{t}<0$, then the distribution of
the increment is computed using the posterior associated with the prior
assigning the largest probability $\frac{1+\epsilon}{2}$ to the positive bias
$\alpha$. It follows that, from the perspective of the worst-case scenario,
the signal structure is less informative the greater is $\epsilon$.
Accordingly, conditional on some learning being optimal, then it must be with
the expectation of a long sampling period that increases in length with
$\epsilon$. A second effect of an increase in $\epsilon$ is that it reduces
the ex ante utility of betting on the ambiguous urn and hence implies that
signals in an increasingly large interval would not change betting preference.
Consequently, a small sample is unlikely to be of value -- only long samples
are useful. Together, these two effects suggest existence of a cutoff value
for $\epsilon$ beyond which no amount of learning is sufficiently attractive
to justify its cost. At the cutoff, here $2\widehat{r}-1$, DM is just
indifferent between stopping and learning for another instant.

There remains the following question for smaller values of $\epsilon$: why is
it never optimal to try learning for a while and then, for some sample
realizations, to stop and bet on the risky urn? The intuition, adapted from
Fudenberg, Strack and Strzalecki (2018), is that this feature is a consequence
of the specification $\mathcal{M}_{0}$ for the set of priors. To see why,
suppose that $Z_{t}$ is small for some positive $t$. A possible
interpretation, particularly for large $t$, is that the true bias is small and
thus that there is little to be gained by continuing to sample -- DM might as
well stop and bet on the risky urn. But this reasoning is excluded when, as in
our specification, DM is certain that the bias is $\pm\alpha$. Then signals
sufficiently near $0$ must be noise and the situation is essentially the same
as it was at the start. Hence, if stopping to bet on the risky urn were
optimal at $t$, it would have been optimal also at time $0 $. This intuition
is suggestive of the likely consequences of generalizing the specification of
$\mathcal{M}_{0}$. Suppose, for example, that $\mathcal{M}_{0}$ is such that
all its priors share a common finite support. We conjecture that then the
predicted incompatibility of learning and betting on the risky urn would be
overturned if the zero bias point is in the common support.

Finally, using the closed-form solution in the theorem, we can give more
concrete expression to the effect of ambiguity on optimal learning. Restrict
attention to values of $\epsilon$ in $[0,2\widehat{r}-1)$, where some learning
is optimal, and denote by $P^{\theta}$ the probability distribution of
$\left(  Z_{t}\right)  $ if $\theta$ is the true bias. Then, by well-known
results regarding hitting times of Brownian motion with drift (Borodin and
Salminen 2015), the mean sample length according to $P^{\theta} $ is%
\begin{equation}
E^{\theta}\tau^{\ast}=\left\{
\begin{array}
[c]{cc}%
\left(  \overline{z}/\sigma\right)  ^{2}\left[  \frac{\tanh\left(
\theta\overline{z}/\sigma^{2}\right)  }{\theta\overline{z}/\sigma^{2}}\right]
& \text{if }\theta\not =0\\
\left(  \overline{z}/\sigma\right)  ^{2} & \text{ if }\theta=0\text{,}%
\end{array}
\right. \label{Etau}%
\end{equation}
which is increasing in $\epsilon$. Note also that $\theta Z_{\tau^{\ast}}>0$
if and only if the bet on red (blue) is chosen on stopping if $\theta>0$
($\theta$ $<0$). Thus the probability, if $\theta\not =0$ is the true bias, of
choosing the "correct" bet on stopping is given by
\[
P^{\theta}\left(  \{\theta Z_{\tau^{\ast}}>0\}\right)  =\frac{1}{1+\exp\left(
-\frac{2\mid\theta\mid}{\sigma^{2}}\overline{z}\right)  }\text{, \ if }%
\theta\not =0\text{,}%
\]
which increases with $\epsilon$. (To prove this equality, apply the optional
stopping theorem to the $P^{\theta}$-martingale $e^{-2\theta Z_{t}/\sigma^{2}%
}$.)

The proof of Theorem \ref{thm-ellsberg} yields a closed-form expression for
the value function associated with the optimal stopping problem. In
particular, the value at time $0$ satisfies{\large \ }(from (\ref{v1}) and
(\ref{v2})),
\begin{equation}
v_{0}-\tfrac{1}{2}=\left\{
\begin{array}
[c]{lc}%
0 & \text{if }\frac{1+\epsilon}{2}\geq\widehat{r}\\
\frac{c\sigma^{2}}{4\alpha^{2}}[\frac{1}{\overline{r}(1-\overline{r})}%
-\frac{4}{(1+\epsilon)(1-\epsilon)}] & \text{if }\frac{1+\epsilon}%
{2}<\widehat{r}\text{.}%
\end{array}
\right. \label{v0}%
\end{equation}
Since the payoff $\frac{1}{2}$ is the best available without learning,
$v_{0}-\frac{1}{2}$ is the value of the learning option. It is positive for
small $\epsilon<2\widehat{r}-1$ and declines continuously to $0$ as $\epsilon$
increases to the switch point. (Note that $\frac{1+\epsilon}{2}=\widehat{r}$
implies both are equal in turn to $\overline{r}$, and hence that $v_{0}$ is
continuous at $\epsilon=2\widehat{r}-1$.) This is consistent with intuition
given above.

As a numerical example, let $\left(  c,\sigma,\alpha\right)  =\left(
.01,1,\frac{1}{8}\right)  $, which gives $.0488$ as the cutoff for $\epsilon$.
Thus learning is rejected if $\epsilon=.05$. For $\epsilon=.04$, however,
$\tau^{\ast}>0$ and$\ E\tau^{\ast}=~.61$ under $P^{\theta=0}$. $\ $Neither of
the values for $\epsilon$ is extreme: in the classic Ellsberg setting (with no
learning), they imply probability equivalents for the bet on red equal to
$.{\small 4875}$ and ${\small .4900}$ for $\epsilon=.05$ and $\epsilon=.04$ respectively.

\subsection{A robust sequential hypothesis test\label{section-test}}

DM samples the signal process $Z$ with the objective of then choosing between
the two statistical hypotheses%
\[
H_{0}:\theta=0\text{ and }H_{1}:\theta=\beta\text{,}%
\]
where $\beta>0$. The novelty relative to Arrow, Blackwell and Girschik (1949)
and Peskir and Shiryaev (2006) is that there is prior ambiguity about the
value of $\theta$ and a robust decision procedure is sought.

The following specialization of the general model is adopted. Let
$\Theta=\{0,\beta\}$. The actions $a_{0}$ and $a_{1}$ are accept $H_{0}$ and
accept $H_{1}$, respectively. A third action is absent because there is no
"outside option" - one of the hypotheses must be chosen. (Formally, one could
include $a_{2}$ and specify its payoff below to be zero, in which case it
would never be chosen.) The set of priors $\mathcal{M}_{0}$ is as given in
(\ref{M0}), corresponding to the probability interval $\left[  \underline{m}%
_{0},\overline{m}_{0}\right]  $ for $\theta=\beta$. Finally, payoffs are given
by%
\begin{align*}
u\left(  a_{0},0\right)   & =u\left(  a_{1},\beta\right)  =a+b\text{, }\\
u\left(  a_{0},\beta\right)   & =b\text{, }u\left(  a_{1},0\right)  =a\text{,}%
\end{align*}
where $a,b>0$. (Payoffs in this context are usually specified in terms of a
loss function that is to be minimized. The loss function $L$ satisfying
$L\left(  a_{0},0\right)  =L\left(  a_{1},\beta\right)  =0$, $L\left(
a_{0},\beta\right)  =a$, and $L\left(  a_{1},0\right)  =b$, gives an
equivalent reformulation.)

There are two differences in specification from the Ellsberg context. First,
there is no counterpart of the risky urn when choosing between hypotheses.
Second, while symmetry between colors is natural in the Ellsberg context,
symmetry between hypotheses is not; thus, $b$ need not equal $a$ and the
probability interval $\left[  \underline{m}_{0},\overline{m}_{0}\right]  $
need not be symmetric about $\frac{1}{2}$.

The optimal stopping problem (\ref{stopX}) admits a closed-form solution. For
perspective, consider first the special Bayesian case ($\mathcal{M}_{0}%
=\{\mu\}$, hence $\mathcal{M}_{t}=\{\mu_{t}\}$, $\mu_{t}\left(  \beta\right)
=m_{t}$). Denote by $\tilde{r}_{B}^{\ell}<\tilde{r}_{B}^{R}$\ the solutions to
(\ref{nonsymmetry-equ}), which in this context simplifies to%
\begin{equation}%
\begin{array}
[c]{l}%
l(\tilde{r}_{B}^{R})-l(\tilde{r}_{B}^{l})=\frac{a+b}{\hat{c}}\\
\frac{1}{\tilde{r}_{B}^{R}\left(  1-\tilde{r}_{B}^{R}\right)  }-\frac
{1}{\tilde{r}_{B}^{l}\left(  1-\tilde{r}_{B}^{l}\right)  }=\frac{b-a}{\hat{c}%
}.
\end{array}
\label{r-tildaB}%
\end{equation}
Then we have the following classical result.

\begin{theorem}
[Peskir and Shiryaev 2006]In the Bayesian case, for any prior probability
$m_{0}$ it is optimal to continue at $t$ if and only if
\begin{equation}
\tilde{r}_{B}^{\ell}<m_{t}<\widetilde{r}_{B}^{R}%
.\label{continue-Bayesian-test}%
\end{equation}
Otherwise, it is optimal to accept $H_{1}$ or $H_{0}$ according as $m_{t}%
\geq\widetilde{r}_{B}^{R}$ or $m_{t}\leq\widetilde{r}_{B}^{\ell}$ respectively.
\end{theorem}

In the model with ambiguity, the cut-off values are $\tilde{r}^{\ell}$\ and
$\tilde{r}^{R}$, $\tilde{r}^{\ell}<\tilde{r}^{R}$, that solve the appropriate
version of (\ref{nonsymmetry-equ}), and we have the following generalization
of the classical result.

\begin{theorem}
\label{thm-test}In the model with ambiguity, it is optimal to stop and accept
$H_{1}$ or $H_{0}$ according as \underline{$m$}$_{t}\geq\widetilde{r}^{R}$ or
$\overline{m}_{t}\leq\widetilde{r}^{\ell}$ respectively. Otherwise, it is
optimal to continue. \newline In addition, if $a=b$, then%
\begin{equation}
\widetilde{r}_{B}^{\ell}<\widetilde{r}^{\ell}\text{ and }\widetilde{r}%
^{R}<\widetilde{r}_{B}^{R}\text{.}\label{r-ineqB}%
\end{equation}

\end{theorem}

Under the assumption of payoff symmetry ($a=b$), the theorem has noteworthy
implications for the relation between the optimal stopping strategies for the
Bayesian and the robustness-seeking DM. (We conjecture that (\ref{r-ineqB}) is
valid even if $a\not =b$, but a proof has escaped us.) If $m_{0}\in\left[
\underline{m}_{0},\overline{m}_{0}\right]  $ refer to a \emph{compatible
Bayesian}. The theorem implies:

\begin{enumerate}
\item If every compatible Bayesian stops and chooses $a_{i}$, then it is
optimal also for DM to stop and choose $a_{i}$, $i=1,2$.

\item If every compatible Bayesian continues, then it may still be optimal for
DM to stop.
\end{enumerate}

\noindent In other words, DM should accept a unanimous recommendation of
compatible Bayesian experts if it is to stop and choose a specific action, but
not necessarily if it is to continue. In this sense, \emph{"sensitivity
analysis" overstates the robustness value of sampling}.

The intuition is clear. Prior ambiguity leads to the signal structure being
perceived as less likely to be informative (seen from the perspective of the
worst-case measure $P^{\ast}$ - see the outline at the start of the proof of
Theorem \ref{thm-general}), even though the signal structure itself is not
ambiguous. In contrast, there is no counterpart given multiple Bayesian agents
- each is confident in beliefs about $\theta$\ and is certain that signal
increments are conditionally i.i.d. Only DM internalizes uncertainty about the
probability law and discounts the benefits of learning accordingly.

\begin{remark}
As is made clear in Theorem \ref{thm-general}, stopping conditions can be
stated equivalently in terms of either the signal process (as in the Ellsberg
model), or posteriors (as here). In the text, we have adopted the formulations
that seem more natural for each particular setting. For example, the use of
posteriors above facilitates comparison with the classical Bayesian result.
\end{remark}

\begin{remark}
Time-consistency in the present context is closely related to the Stopping
Rule Principle -- \ that the stopping rule should have no effect on what is
inferred from observed data and hence on the decision taken after stopping
(Berger 1985). It is well-known that: (i) conventional frequentist methods,
based on ex ante fixed sample size significance levels, violate this Principle
and permit the analyst to sample to a foregone conclusion when data-dependent
stopping rules are permitted; and (ii) Bayesian posterior odds analysis
satisfies the Principle. Kadane, Schervish and Seidenfeld (1996) point to the
law of iterated expectations as responsible for excluding foregone conclusions
(if the prior is countably additive). Equation (\ref{LIE}) is a nonlinear
counterpart that we suspect plays a similar role in our model (though details
are beyond the scope of this paper).
\end{remark}

\section{A more general theorem\label{section-generalthm}}

In order to condense notation, we write $u_{ij}$ in place of $u\left(
a_{i},\theta_{j}\right)  $, $i,j=0,1$.

Theorem \ref{thm-general} below describes the solution to the optimal stopping
problem in \S 3.1 assuming \emph{either} payoff symmetry ($u_{01}=u_{10}$)
\emph{or} no risky option ($u_{2}\leq\min\{u_{10},u_{01}\}$). Payoff symmetry
is satisfied in Theorem \ref{thm-ellsberg}, but the latter assumes more,
specifically ex ante indifference between $a_{0}$ and $a_{1}$ (\underline{$m$%
}$_{0}+\overline{m}_{0}=1$) and $u_{2}=\frac{1}{2}(u_{00}+u_{10})$. Thus it is
extended below by Theorem \ref{thm-general}(a). The assumption of no risky
option is the crucial element in the hypothesis testing example, and the
corresponding optimal stopping problem is isomorphic to that in part (b) of
Theorem \ref{thm-general}.

Both $\overline{m}_{t}$ and $\underline{m}_{t}$ defined in (\ref{mt}) are
increasing functions of $\varphi(t,z_{t})$. It follows that there exists a
unique pair of probabilities \underline{$\pi$} and $\overline{\pi}$ and a
unique (deterministic) signal realization trajectory $(\widetilde{z}_{t})$
satisfying, for every $t$,
\[
\underline{\pi}=\underline{m}_{t}(\widetilde{z}_{t})\text{, }\overline{\pi
}=\overline{m}_{t}(\widetilde{z}_{t})\text{, \ \ and}%
\]%
\[
\underline{\pi}u_{11}+\left(  1-\underline{\pi}\right)  u_{10}=\overline{\pi
}u_{01}+\left(  1-\overline{\pi}\right)  u_{00}\text{.}%
\]
For example, $\widetilde{z}_{0}=0$, $\underline{\pi}=\underline{m}_{0}$ and
$\overline{\pi}=\overline{m}_{0}$ if and only if $a_{0}$ and $a_{1}$ are
indifferent ex ante. More generally, $a_{0}$ and $a_{1}$ are indifferent
conditional on the signal $\widetilde{z}_{t}$ at $t$ and $a_{0}$ ($a_{1}$) is
preferred at $t$ if $Z_{t}<\left(  >\right)  \widetilde{z}_{t}$.

Normalize the cost of learning to $\widehat{c}$, $\hat{c}=2c\sigma^{2}%
/(\theta_{1}-\theta_{0})^{2}$.

Optimal stopping strategies will be described in terms of several critical
values, that are, in turn, defined using the functions $l$ and $\widetilde{l}
$: For all $r$ in $\left(  0,1\right)  $,%
\begin{align*}
l(r)  & =2\log(\frac{r}{1-r})-\frac{1}{r}+\frac{1}{1-r}\\
\tilde{l}(r)  & =\log(\frac{r}{1-r})+\frac{r}{1-r}\text{.}%
\end{align*}

Let $(r_{1}^{R},r_{2}^{R})$, $(r_{1}^{l},r_{2}^{l})$, $\left(  r^{R}%
,r^{l}\right)  $ and $\left(  \tilde{r}^{R},\tilde{r}^{l}\right)  $ solve the
following equations respectively:
\begin{equation}%
\begin{array}
[c]{rl}%
l(r_{2}^{R})-l(r_{1}^{R}) & =\frac{u_{11}-u_{10}}{\hat{c}}\\
\tilde{l}(r_{2}^{R})-\tilde{l}(r_{1}^{R}) & =\frac{u_{2}-u_{10}}{\hat{c}%
}\text{,}%
\end{array}
\label{equ1-critical point-1-R}%
\end{equation}%
\begin{equation}%
\begin{array}
[c]{rl}%
l(r_{2}^{l})-l(r_{1}^{l}) & =-\frac{u_{00}-u_{01}}{\hat{c}}\\
\tilde{l}(r_{2}^{l})-\tilde{l}(r_{1}^{l}) & =\frac{u_{2}-u_{00}}{\hat{c}%
}\text{,}%
\end{array}
\label{equ1-critical point-1-L}%
\end{equation}%
\begin{equation}
\left.
\begin{array}
[c]{rl}%
l(r^{R})-l(\underline{\pi}) & =\frac{u_{11}-u_{10}}{\hat{c}}\\
l(r^{l})-l(\overline{\pi}) & =-\frac{u_{00}-u_{01}}{\hat{c}}\text{,}%
\end{array}
\right. \label{equ1-critical point-2}%
\end{equation}%
\begin{equation}%
\begin{array}
[c]{l}%
l(\tilde{r}^{R})-l(\underline{\pi})=l(\tilde{r}^{l})-l(\overline{\pi}%
)+\frac{u_{11}-u_{10}+u_{00}-u_{01}}{\hat{c}}\\
\tilde{l}(\widetilde{r}^{R})-\tilde{l}(\underline{\pi})-\underline{\pi}\left(
l(\widetilde{r}^{R})-l(\underline{\pi})\right)  =\\
\widetilde{l}(\widetilde{r}^{l})-\widetilde{l}(\overline{\pi})-\overline{\pi
}\left(  l(\widetilde{r}^{l})-l(\overline{\pi})\right)  .
\end{array}
\label{nonsymmetry-equ}%
\end{equation}
\bigskip(The latter reduces to (\ref{equ1-critical point-2}) if payoff
symmetry is satisfied.)

Define
\begin{equation}
u_{2}^{\ast\ast}=\frac{\hat{c}}{2}[\frac{1}{r^{l}(1-r^{l})}-\frac{1}%
{\overline{\pi}(1-\overline{\pi})}]+\frac{u_{00}-u_{01}}{2}.\label{u2**}%
\end{equation}

Besides the existence and uniqueness assertions, the next lemma proves a
number of properties that are important for the optimal stopping theorem to follow.

\begin{lemma}
\label{lemma-r}There exist unique solutions to (\ref{equ1-critical point-2})
and (\ref{nonsymmetry-equ}), and the solutions to the latter satisfy
\begin{equation}
\tilde{r}^{l}<\overline{\pi}\text{, }\tilde{r}^{R}>\underline{\pi}%
\text{.}\label{rtilda}%
\end{equation}
If $u_{2}\geq u_{2}^{\ast\ast}$, then there exist unique solutions also to
(\ref{equ1-critical point-1-R}) and (\ref{equ1-critical point-1-L}), and the
solutions satisfy
\[
r_{2}^{l}<r_{1}^{l}\text{, }r_{1}^{R}<r_{2}^{R}\text{, }\underline{\pi}%
<r^{R}\text{, }r^{l}<\overline{\pi}\text{.}%
\]
If payoff symmetry is also satisfied, then:
\begin{equation}
\underline{\pi}+\overline{\pi}=1=r^{l}+r^{R}\text{, \ and}\label{rpi1}%
\end{equation}
{\tiny \ }%
\begin{equation}
r_{1}^{l}\leq\overline{\pi}~\Longleftrightarrow r_{1}^{R}\geq\underline{\pi
}~\Longleftrightarrow u_{2}\geq u_{2}^{\ast\ast}\text{.}\label{2equivalences}%
\end{equation}
\noindent
\end{lemma}

Define%
\begin{align*}
\overline{f}(t,r)  & =\frac{\theta_{1}+\theta_{0}}{2}t+\frac{\sigma^{2}%
}{\theta_{1}-\theta_{0}}\log(\frac{1-\overline{m}_{0}}{\overline{m}_{0}}%
\frac{r}{1-r})\\
\underline{f}(t,r)  & =\frac{\theta_{1}+\theta_{0}}{2}t+\frac{\sigma^{2}%
}{\theta_{1}-\theta_{0}}\log(\frac{1-\underline{m}_{0}}{\underline{m}_{0}%
}\frac{r}{1-r})\text{.}%
\end{align*}
Then $\underline{m}_{t}\left(  \underline{f}(t,r)\right)  =r=\overline{m}%
_{t}\left(  \overline{f}(t,r)\right)  $, and, for any $r_{1}$ and $r_{2}$,
\begin{align}
\overline{f}(t,r_{1})  & \leq\widetilde{z}_{t}\Longleftrightarrow r_{1}%
\leq\overline{\pi}\label{f-ineq}\\
\underline{f}(t,r_{2})  & \geq\widetilde{z}_{t}\Longleftrightarrow r_{2}%
\geq\underline{\pi}\text{.}\nonumber
\end{align}
\noindent Finally, define three stopping times:
\begin{align*}
\tau_{0}  & \equiv\min\{t\geq0:Z_{t}\leq\overline{f}(t,r_{2}^{l})\}\\
& =\min\{t\geq0:\text{ }\overline{m}_{t}\leq r_{2}^{l}\}\text{,}%
\end{align*}%
\begin{align*}
\tau_{1}  & \equiv\min\{t\geq0:Z_{t}\geq\underline{f}(t,r_{2}^{R})\}\\
& =\min\{t\geq0:\underline{m}_{t}\geq r_{2}^{R}\}\text{, and}%
\end{align*}%
\begin{align*}
\tau_{2}  & \equiv\min\{t\geq0:\overline{f}(t,r_{1}^{l})\leq Z_{t}%
\leq\underline{f}(t,r_{1}^{R})\}\\
& =\min\{t\geq0:\overline{m}_{t}\geq r_{1}^{l}\text{ and }\underline{m}%
_{t}\leq r_{1}^{R}\}\text{.}%
\end{align*}

\begin{theorem}
\label{thm-general}(a) Assume payoff symmetry ($u_{01}=u_{10}$).

(a.i) If $r_{1}^{l}\leq\overline{\pi}$, then the optimal stopping time
$\tau^{\ast}$ is given by%
\[
\tau^{\ast}=\min\{\tau_{i}:i=0,1,2\}\text{.}%
\]
Moreover, if $\tau^{\ast}=\tau_{i}$, then $a_{i}$ is optimal on stopping. In
particular, if there is ex ante indifference between $a_{0}$ and $a_{1}$
($\underline{\pi}=\underline{m}_{0}$ and $\overline{\pi}=\overline{m}_{0}$),
then $\tau^{\ast}=0$ and $a_{2}$ is chosen.

(a.ii) If $r_{1}^{l}>\overline{\pi}$, then%
\begin{align*}
\tau^{\ast}  & =\min\{t\geq0:\text{ }Z_{t}\leq\overline{f}(t,r^{l})\text{ or
}Z_{t}\geq\underline{f}(t,r^{R})\}\\
& =\min\{t\geq0:\text{ }\overline{m}_{t}\leq r^{l}\text{ or }\underline{m}%
_{t}\geq r^{R}\}\text{.}%
\end{align*}
Moreover, $a_{0}$ is optimal on stopping if $Z_{\tau^{\ast}}\leq\overline
{f}(\tau^{\ast},r^{l})~$(equivalently if $\overline{m}_{\tau^{\ast}}\leq
r^{l}$), $a_{1}$ is optimal if $Z_{\tau^{\ast}}\geq\overline{f}(\tau^{\ast
},r^{R})$ (equivalently if $\underline{m}_{\tau^{\ast}}\geq r^{R}$), and
$a_{2}$ is never optimal.

(b) Assume $u_{2}\leq\min\{u_{10},u_{01}\}$. Then
\begin{align*}
\tau^{\ast}  & =\min\{t\geq0:\text{ }Z_{t}\leq\overline{f}(t,\tilde{r}%
^{l})\text{ or }Z_{t}\geq\underline{f}(t,\tilde{r}^{R})\}\\
& =\min\{t\geq0:\text{ }\overline{m}_{t}\leq\tilde{r}^{l}\text{ or
}\underline{m}_{t}\geq\tilde{r}^{R}\}\text{.}%
\end{align*}
Moreover, $a_{0}$\ is optimal on stopping if $Z_{\tau^{\ast}}\leq\overline
{f}(\tau^{\ast},\tilde{r}^{l})~$(equivalently if $\overline{m}_{\tau^{\ast}%
}\leq\tilde{r}^{l}$), $a_{1}$\ is optimal if $Z_{\tau^{\ast}}\geq
\underline{f}(\tau^{\ast},\tilde{r}^{R})$\ (equivalently if $\underline{m}%
_{\tau^{\ast}}\geq\tilde{r}^{R}$), and $a_{2}$\ is never optimal.
\end{theorem}

In (a), the distinction between the two subcases depends on the relative
magnitudes of $r_{1}^{l}$ and $\overline{\pi}$. From
(\ref{equ1-critical point-1-L}) it follows that $r_{1}^{l}$ falls as $u_{2}$
increases, while $\overline{\pi}$ does not depend on $u_{2}$. Therefore, (a.i)
applies if the payoff $u_{2}$ to the unambiguous default is sufficiently
large. The other factor leading to (a.i) is large $\overline{\pi}$,
equivalently (by (\ref{rpi1})) small \underline{$\pi$}, which is supported by
$\overline{m}_{0}$ large and \underline{$m$}$_{0}$ small. Thus, (a.i) is
supported also by large prior ambiguity.

In (a.i), $\tau^{\ast}=0$ if either $\overline{m}_{0}\leq r_{2}^{l}$ (prior
beliefs are strongly biased towards $\theta_{0}$ and hence $a_{0}$ is chosen
immediately), or \underline{$m$}$_{0}\geq r_{2}^{R}$ (prior beliefs are
strongly biased towards $\theta_{1}$ and hence $a_{1}$ is chosen), or
$\overline{m}_{0}\geq r_{1}^{l}$ and $\underline{m}_{0}\leq r_{1}^{R}$ (the
worst-case probabilities of both $\theta_{0}$ and $\theta_{1}$ are both
sufficiently low that neither $a_{0}$ nor $a_{1}$ are attractive enough to
justify the cost of sampling and hence $a_{2}$ is chosen). That leaves
continuation being optimal at time 0 if and only if prior beliefs are
"intermediate" in the sense that%
\begin{align*}
\text{either}\text{: }  & [r_{2}^{l}<\overline{m}_{0}<r_{1}^{l}]\text{ and
}\underline{m}_{0}<r_{2}^{R}\text{,}\\
\text{or}\text{: }  & [r_{1}^{R}<\underline{m}_{0}<r_{2}^{R}]\text{ and
}\overline{m}_{0}>r_{2}^{l}]\text{.}%
\end{align*}
This continuation region could be empty. Since learning is only about the
payoffs to $a_{0}$\textbf{\ }and $a_{1}$, the situation at time $0$ that is
least favorable to learning is where there is ex ante indifference between
$a_{0}$ and $a_{1}$ -- then a long and hence costly sample would likely be
needed to modify the ex ante ranking of actions. In this case, therefore, it
is optimal to reject learning and choose $a_{2}$, as in Theorem
\ref{thm-ellsberg}. However, if, for example, $a_{1}$ is strictly preferred
initially, then an incentive to learn is that a relatively short interval of
sampling may be enough to decide between $a_{1}$ and $a_{2}$. In addition, if
$\underline{m}_{0}$ is sufficiently large, say near $1$, then near certainty
that $\theta=\theta_{1}$ can lead to rejection of learning and the immediate
choice of $a_{1}$, rather than of $a_{2}$ as in the Ellsberg context.

In (a.ii), $\tau^{\ast}=0$ iff $\left[  \underline{m}_{0},\overline{m}%
_{0}\right]  $ is disjoint from $(r^{\ell},r^{R})$. Notably, the default
action is not chosen regardless of when sampling stops. Its payoff $u_{2}$ is
too low (from (\ref{2equivalences}), $u_{2}<u_{2}^{\ast\ast}$) compared to the
expected payoff of choosing $a_{0}$ or $a_{1}$, possibly after some learning.
Moreover, even given some learning, it is not optimal to choose $a_{2}$
regardless of the realized sample, as explained in discussion of Theorem
\ref{thm-ellsberg}. Under ex ante indifference, Lemma \ref{lemma-r} implies
that $\tau^{\ast}>0$ in (a.ii). Combined with (a.i), we see that if there is
ex ante indifference between $a_{0}$ and $a_{1}$, \emph{then }$a_{2}%
$\emph{\ is chosen if and only if there is no learning}, thus generalizing the
result in the Ellsberg model. \textbf{(}The latter also assumes $u_{2}%
=\frac{1}{2}(u_{00}+u_{10})$, which we see here is not needed for the
preceding conclusion.)

Finally, consider (b), where the payoff to the unambiguous action is so low
that it would never be chosen, regardless of prior beliefs and even in the
absence of the option to learn. The optimal strategy is similar to that in
(a.ii) in form and interpretation - only the critical values may differ to
reflect the different assumptions about payoffs. Another comment about (b) is
that when $\overline{m}_{0}=\underline{m}_{0}$, then $\overline{\pi
}=\underline{\pi}$\ and the equations (\ref{nonsymmetry-equ}) defining the
critical values $\tilde{r}^{R}$ and $\tilde{r}^{l}$ become%
\[%
\begin{array}
[c]{l}%
l(\tilde{r}^{R})-l(\tilde{r}^{l})=\frac{u_{11}-u_{10}+u_{00}-u_{01}}{\hat{c}%
}\\
\tilde{l}(\tilde{r}^{R})-\tilde{l}(\tilde{r}^{l})=\frac{u_{00}-u_{10}}{\hat
{c}}\text{,}%
\end{array}
\]
which are equations (21.1.14) and (21.1.15) in Peskir and Shiryaev (2006).

\bigskip

Proof of the theorem is provided in the e-companion. Here we comment briefly
on the proof strategy.

The strategy is to: (i) guess the $P^{\ast}$ in $\mathcal{P}_{0}$ that is the
worst-case scenario; (ii) solve the classical optimal stopping problem given
the single prior $P^{\ast}$; (iii) show that the value function derived in (2)
is also the value function for our problem (\ref{stopX}); and (4) use the
value function to derive $\tau^{\ast}$.

The intuition for the conjectured $P^{\ast}$ was given in \S 3.2 for the
Ellsberg context. In this more general context, it extends to the conjecture
that $P^{\ast}$ should make $P^{\ast}(\{dZ_{t}>0\}\mid Z_{t}>\widetilde{z}%
_{t})$\ and $P^{\ast}(\{dZ_{t}<0\}\mid Z_{t}<\widetilde{z}_{t})${\small \ }as
small as possible, by using \underline{$m$}$_{t}$ when $Z_{t}>\widetilde{z}%
_{t}$ and $\overline{m}_{t}$ when $Z_{t}<\widetilde{z}_{t}$. (See
(\ref{P*-density}) for the precise definition of $P^{\ast}$.) The search for
the value function $v$ begins with the HJB equation which yields its
functional form up to some constants to be determined by smooth contact
conditions between $v$ and the payoff function $X$ (see Peskir and Shiryaev
(2006) for this free-boundary approach to analysing optimal stopping
problems). A new ingredient relative to existing models stems from the nature
of $P^{\ast}$, specifically from the fact that the relevant posterior
probability at $t$ switches between \underline{$m$}$_{t}$ and $\overline
{m}_{t}$ as described, implying that the form of the value function differs
between the regions $Z_{t}>\widetilde{z}_{t}$ and $Z_{t}<\widetilde{z}_{t}$.
Thus, in addition to ensuring a smooth contact at stopping points, one must
also be concerned with the smooth connection at $\widetilde{z}_{t}$.

We elaborate on the latter point in order to highlight the technical novelty
that arises from ambiguity. For concreteness consider (a.ii), where $a_{2}$ is
never chosen. Let $y$ denote a posterior probability, computed using
\underline{$m$}$_{0}$ or $\overline{m}_{0}$, depending on the sub-domain, and
let $V^{R}(y):[\underline{\pi},1]\rightarrow\lbrack0,+\infty)$ and
$V^{l}(y):[0,\overline{\pi}]\rightarrow\lbrack0,+\infty)$ denote corresponding
candidates for the value in the indicated regions. Then the variational
inequality and smooth contacts lead to the following free-boundary
differential equation, in which $r^{R}\in(\underline{\pi},1]$ and $r^{l}%
\in\lbrack0,\overline{\pi})$ are also unknowns to be determined:
\begin{equation}
\left\{
\begin{array}
[c]{rl}%
V_{yy}^{R}(y) & =\hat{c}\frac{1}{y^{2}(1-y)^{2}},\;\ y\in(\underline{\pi
},r^{R})\\
V^{R}(r^{R}) & =(u_{11}-u_{10})r^{R}+u_{10}\\
V_{y}^{R}(r^{R}) & =(u_{11}-u_{10})\\
V_{yy}^{l}(y) & =\hat{c}\frac{1}{y^{2}(1-y)^{2}},\;\ y\in(r^{l},\overline{\pi
})\\
V^{l}(r^{l}) & =-(u_{00}-u_{01})r^{l}+u_{00}\\
V_{y}^{l}(r^{l}) & =-(u_{00}-u_{01})\text{,}%
\end{array}
\right. \label{free-equation}%
\end{equation}
and the (new) smooth contact conditions due to ambiguity ($\underline{\pi
}<\overline{\pi}$):%
\begin{equation}
\left\{
\begin{array}
[c]{rl}%
V^{R}(\underline{\pi}) & =V^{l}(\overline{\pi}),\\
V_{y}^{R}(\underline{\pi}) & =V_{y}^{l}(\overline{\pi}).
\end{array}
\right. \label{switching smooth}%
\end{equation}
In (a.ii), payoff symmetry leads to the simplification $V_{y}^{R}%
(\underline{\pi})=V_{y}^{l}(\overline{\pi})=0$, which leads to
(\ref{equ1-critical point-2}) becoming two separated equations. However, in
(b), the connection is not trivial.

\section{Proofs}

\subsection{Proof of Theorem \ref{thm-general}\label{app-generalproof}}

Below "almost surely" qualifications should be understood, even where not
stated explicitly, and as defined relative to any measure in $\mathcal{P}_{0}
$.

To compute the payoff $X_{t}$ defined in (\ref{Xt}), note that
\begin{align*}
\min_{\mu\in\mathcal{M}_{t}}\int u\left(  a_{0},\theta\right)  d\mu &
=(u_{00}-u_{01})(1-\overline{m}_{t})+u_{01},\\
\min_{\mu\in\mathcal{M}_{t}}\int u\left(  a_{1},\theta\right)  d\mu &
=(u_{11}-u_{10})\underline{m}_{t}+u_{10}.
\end{align*}
There is a critical level of $u_{2}$, denoted $u_{2}^{\ast}$,
\[
u_{2}^{\ast}=\frac{u_{11}u_{00}-u_{10}u_{01}}{u_{00}+u_{11}-u_{01}-u_{10}%
}\text{.}%
\]
If $u_{2}\leq u_{2}^{\ast}$, then $X_{t}=$
\[
\left\{
\begin{array}
[c]{lc}%
(u_{00}-u_{01})(1-\overline{m}_{t})+u_{01} & \text{if }\overline{m}%
_{t}<\overline{\pi}\\
(u_{11}-u_{10})\underline{m}_{t}+u_{10} & \text{if }\underline{m}_{t}%
\geq\underline{\pi}\text{.}%
\end{array}
\right.
\]
Accordingly, the default action $a_{2}$ is not optimal at any $t$, and $a_{0}
$ ($a_{1}$) is optimal conditional on stopping at $t$ if $\overline{m}%
_{t}<\overline{\pi}$ ($\underline{m}_{t}\geq\underline{\pi}$). If $u_{2}%
>u_{2}^{\ast}$, then $X_{t}=$
\[
\left\{
\begin{array}
[c]{ll}%
(u_{00}-u_{01})(1-\overline{m}_{t})+u_{01} & \text{if }\overline{m}_{t}%
<\frac{u_{00}-u_{2}}{u_{00}-u_{01}}\\
(u_{11}-u_{10})\underline{m}_{t}+u_{10} & \text{if }\underline{m}_{t}\geq
\frac{u_{2}-u_{10}}{u_{11}-u_{10}}\\
u_{2} & \text{otherwise,}%
\end{array}
\right.
\]
reflecting the conditional optimality of $a_{0}$, $a_{1}$ and $a_{2}$
respectively in the three indicated regions.

As in \S 2, for any $\mu\in\mathcal{M}_{0}$, $\mu_{t}$ denotes its Bayesian
posterior at $t$ and $\widehat{\theta}_{t}^{\mu}=\int\theta d\mu_{t}$ is the
corresponding posterior estimate of $\theta$. The two extreme measures
$\mu=\overline{\mu}$, \underline{$\mu$}, are defined by
\[
\overline{\mu}_{t}\left(  \theta_{1}\right)  =\overline{m}_{t}\text{ and
}\underline{\mu}_{t}\left(  \theta_{1}\right)  =\underline{m}_{t}\text{,}%
\]
and yield the estimates $\hat{\theta}_{t}^{\overline{\mu}}$ and $\hat{\theta
}_{t}^{\underline{\mu}}$ respectively. Let $P^{\ast}$ be the probability
measure in $\mathcal{P}_{0}$ which has density generator process $\left(
\eta_{t}\right)  $,%
\begin{equation}
-\eta_{t}=(\hat{\theta}_{t}^{\overline{\mu}}/\sigma)\boldsymbol{1}_{Z_{t}%
\leq\widetilde{z}_{t}}+(\hat{\theta}_{t}^{\underline{\mu}}/\sigma
)\boldsymbol{1}_{Z_{t}>\widetilde{z}_{t}}\text{.}\label{P*-density}%
\end{equation}
It will be shown that $P^{\ast}$ is the worst-case scenario in $\mathcal{P}%
_{0}$.

\bigskip

\noindent\textbf{Proof of (a.ii)}: Consider the classical optimal stopping
problem under $P^{\ast}$,
\begin{equation}
\underset{\tau}{\max}E_{P^{\ast}}[X_{\tau}-c\tau]\text{.}%
\label{stopX-classical}%
\end{equation}

Define $g_{1}$ and $g_{2}$ by, for $0<y<1$, $i=1,2$,
\begin{equation}
g_{i}(y;C_{2i-1},C_{2i})=\hat{c}(2y-1)\log(\frac{y}{1-y})+C_{2i-1}%
y+C_{2i}\text{,}\;\label{g1g2}%
\end{equation}
where the constants $C_{i}$ ($i=1$, $2$, $3$, $4$) are determined by
smooth-contact conditions.

We conjecture that the value function for (\ref{stopX-classical}) has the
form: $v(t,z)=$%

\begin{equation}
\left\{
\begin{array}
[c]{cc}%
(u_{00}-u_{01})(1-\overline{m}_{t}\left(  z\right)  )+u_{01} & \text{if
}z<\overline{f}(t,r^{l})\\
g_{1}(\overline{m}_{t}\left(  z\right)  ;C_{1},C_{2}) & \text{if }\overline
{f}(t,r^{l})\leq z<\widetilde{z}_{t}\\
g_{2}(\underline{m}_{t}\left(  z\right)  ;C_{3},C_{4}) & \text{if
}\widetilde{z}_{t}\leq z<\underline{f}(t,r^{R})\\
(u_{11}-u_{10})\underline{m}_{t}\left(  z\right)  +u_{10} & \text{if
}\underline{f}(t,r^{R})\leq z,
\end{array}
\right. \label{v1}%
\end{equation}
where
\begin{align*}
C_{1}  & =-\hat{c}\ell(\overline{\pi})\text{, \ }C_{3}=-\hat{c}\ell
(\underline{\pi})\\
C_{2}  & =(u_{00}-u_{01})(1-r^{l})+u_{01}\\
& -\hat{c}[(2r^{l}-1)\log(\frac{r^{l}}{1-r^{l}})-\ell(\overline{\pi})r^{l}]\\
\text{\ }C_{4}  & =(u_{11}-u_{10})r^{R}+u_{10}\\
& -\hat{c}[(2r^{R}-1)\log(\frac{r^{R}}{1-r^{R}})-\ell(\underline{\pi}%
)r^{R}]\text{.}%
\end{align*}
\textbf{\ }(Note that the cut-off value $u_{2}^{\ast\ast}$ defined in
(\ref{u2**}) satisfies $u_{2}^{\ast\ast}=g_{1}(\overline{\pi};C_{1}%
,C_{2})=g_{2}(\underline{\pi};C_{3},C_{4})=v(t,\widetilde{z}_{t})$.)

\begin{lemma}
\label{lem-classical} $v$ is the value function of the classical optimal
stopping problem (\ref{stopX-classical}), i.e., for any $t\geq0$,
\[
v(t,z)=\underset{\tau\geq t}{\max}E_{P^{\ast}}[X_{\tau-t}-c(\tau-t)\mid
Z_{t}=z]\text{.}%
\]
Further, $v$ satisfies the HJB equation%
\begin{equation}
\max\{X(t,z)-v(t,z),-c+v_{t}(t,z)+\frac{1}{2}\sigma^{2}v_{zz}(z)+f(t,z)v_{z}%
(t,z)\}=0\text{,}\label{HJB}%
\end{equation}
where $f(t,z)\equiv$%
\begin{equation}
\lbrack\theta_{1}-\frac{\theta_{1}-\theta_{0}}{1+\frac{\overline{m}_{0}%
}{1-\overline{m}_{0}}\varphi(t,z)}]1_{\{z<\widetilde{z}_{t}\}}+[\theta
_{1}-\frac{\theta_{1}-\theta_{0}}{1+\frac{\underline{m}_{0}}{1-\underline{m}%
_{0}}\varphi(t,z)}]1_{\{z\geq\widetilde{z}_{t}\}}\text{.}%
\label{density function}%
\end{equation}
Finally, $v$ also satisfies, $\forall z\in(\overline{f}(t,r^{l}),\underline{f}%
(t,r^{R}))$,%
\begin{equation}
-c+v(t,z)+\frac{1}{2}\sigma^{2}v_{zz}(z)+f(t,z)v_{z}(t,z)=0.\label{vxx=0}%
\end{equation}

\end{lemma}

\noindent For the proof, first verify that $v$ satisfies the HJB equation
(\ref{HJB}), and then apply El Karoui et al. (1997, Theorems 8.5, 8.6).
Alternatively, a proof can be constructed along the lines of Peskir and
Shiryaev (2006, Ch. 6).

Next prove that $v$ is the value function of the (nonclassical) optimal
stopping problem (\ref{stopX}) (solving the HJB equation is not sufficient to
imply this). We consider only $t=0$ and prove
\[
v(0,z)=\underset{\tau\geq0}{\max}\underset{P\in\mathcal{P}_{0}}{\min}%
E_{P}[X(Z_{\tau})-c\tau]\text{.}%
\]
By Lemma \ref{lem-classical},
\[
v(0,z)=\underset{\tau\geq0}{\max}E_{P^{\ast}}[X(Z_{\tau})-c\tau]\geq
\underset{\tau\geq0}{\max}\underset{P\in\mathcal{P}_{0}}{\min}E_{P}[X(Z_{\tau
})-c\tau]\text{.}%
\]

To prove the opposite inequality, consider the stopping time%
\[
\tau^{\ast}=\inf\{t\geq0:\text{ }Z_{t}\leq\overline{f}(t,r^{l})\text{ or
}Z_{t}\geq\underline{f}(t,r^{R})\}.
\]
For $t\leq\tau^{\ast}$, by Ito's formula, (\ref{HJB}), and (\ref{vxx=0}),
$dv(t,Z_{t})=$
\begin{align}
& [v_{t}(t,Z_{t})+\frac{1}{2}\sigma^{2}v_{zz}(t,Z_{t})]dt+v_{z}(t,Z_{t}%
)dZ_{t}\label{use-HJB}\\
& =[c-f(t,Z_{t})v_{z}(t,Z_{t})]dt+v_{z}(t,Z_{t})dZ_{t}\nonumber\\
& =[c-f(t,Z_{t})v_{z}(t,Z_{t})]dt+v_{z}(t,Z_{t})dZ_{t}\text{.}\nonumber
\end{align}
Each $P=P^{\eta}\in\mathcal{P}_{0}$ corresponds to a density generator process
$\left(  \eta_{t}\right)  $, and $(W_{t}^{\eta})$ is a Brownian motion under
$P^{\eta}$, where
\[
W_{t}^{\eta}=\frac{1}{\sigma}Z_{t}+\frac{1}{\sigma}\int\nolimits_{0}^{t}%
\tilde{f}(s,Z_{s},\eta_{s})ds\text{, and}%
\]%
\[
\tilde{f}(t,Z_{t},\eta_{t})=[\theta_{1}-\frac{\theta_{1}-\theta_{0}}%
{1+\frac{\eta_{t}}{1-\eta_{t}}\varphi(t,Z_{t})}]\text{.}%
\]
Therefore, $dv(t,Z_{t})=$%
\begin{equation}
\lbrack c+\left(  \tilde{f}(t,Z_{t},\eta_{t})-f(t,Z_{t})\right)  v_{z}%
(t,Z_{t})]dt+\sigma v_{z}(t,Z_{t})dW_{t}^{\eta}.\nonumber
\end{equation}
Note that $\left(  \tilde{f}(t,Z_{t},\eta_{t})-f(t,Z_{t})\right)  v_{z}%
(Z_{t})\geq0$. (Suppose $Z_{t}<\widetilde{z}_{t}$. Then $v_{z}(Z_{t})\leq0$
and $\tilde{f}(t,Z_{t},\eta_{t})-f(t,Z_{t})\leq0$, the latter because
$[\theta_{1}-\frac{\theta_{1}-\theta_{0}}{1+\frac{m}{1-m}\varphi(t,z)}]$ is
increasing in $m$. Argue similarly for $Z_{t}<\widetilde{z}_{t}$.) Take
expectation above under $P^{\eta}$ to obtain%
\begin{align*}
v(0,z)  & \leq E_{P^{\eta}}[v(\tau^{\ast},Z_{\tau^{\ast}})-c\tau^{\ast}]\\
& =E_{P^{\eta}}[X_{\tau^{\ast}}-c\tau^{\ast}].
\end{align*}
The above inequality is due to{\large \ }%
\[
E_{P^{\eta}}[%
{\displaystyle\int\nolimits_{0}^{\tau^{\ast}}}
\sigma v_{z}(t,Z_{t})dW_{t}^{\eta}]=0\text{,}%
\]
which is guaranteed by
\begin{equation}
\underset{P\in\mathcal{P}_{0}}{\max}E_{P}[\tau^{\ast}]<\infty\text{;}%
\label{tau*}%
\end{equation}
see Peskir and Shiryaev (2006, Theorem 21.1) for the classical case. In our
setting, (\ref{tau*}) is implied by the boundedness of $X_{t}$ because:
\begin{align*}
-\infty & <\max_{\tau\geq0}\min_{P\in\mathcal{P}_{0}}E_{P}\left(  X_{\tau
}-c\tau\right)  =\max_{\tau\geq0}[-\underset{P\in\mathcal{P}_{0}}{\max}%
E_{P}\left(  c\tau-X_{\tau}\right)  ]\\
& \leq\max_{\tau\geq0}[\underset{P\in\mathcal{P}_{0}}{\max}E_{P}\left(
X_{\tau}\right)  -\underset{P\in\mathcal{P}_{0}}{\max}E_{P}\left(
c\tau\right)  ]\Longrightarrow\underset{P\in\mathcal{P}_{0}}{\max}E_{P}%
[\tau^{\ast}]<\infty\text{.}%
\end{align*}

Finally, because $P^{\eta}$ can be any measure in $\mathcal{P}_{0}$, deduce
that
\begin{align*}
v(0,z)  & \leq\underset{P\in\mathcal{P}_{0}}{\min}E_{P}[X_{\tau^{\ast}}%
-c\tau^{\ast}]\\
& \leq\underset{\tau\geq0}{\max}\underset{P\in\mathcal{P}_{0}}{\min}%
E_{P}[X_{\tau}-c\tau]\text{.}%
\end{align*}
Conclude that $v$ is the value function for our optimal stopping problem and
that $\tau^{\ast}$ is the optimal stopping time.$\hfill$

\begin{remark}
The preceding implies that $P^{\ast}$\emph{\ }is indeed the minimizing measure
because the minimax property is satisfied:%
\begin{align*}
\max_{\tau\geq0}E_{P^{\ast}}X\left(  Z_{\tau}\right)   & =\max_{\tau\geq0}%
\min_{P\in\mathcal{P}_{0}}E_{P}X\left(  Z_{\tau}\right)  \leq\\
\min_{P\in\mathcal{P}_{0}}\max_{\tau\geq0}E_{P}X\left(  Z_{\tau}\right)   &
\leq\max_{\tau\geq0}E_{P^{\ast}}X\left(  Z_{\tau}\right)  \Longrightarrow\\
\min_{P\in\mathcal{P}_{0}}\max_{\tau\geq0}E_{P}X\left(  Z_{\tau}\right)   &
=\max_{\tau\geq0}\min_{P\in\mathcal{P}_{0}}E_{P}X\left(  Z_{\tau}\right)
\text{.}%
\end{align*}

\end{remark}

$\hfill$

\noindent\noindent\textbf{Proof of (a.i):}\ The proof is similar to that of
(a.ii). The only difference is that the value function $v$ is given by
$v(t,z)=$
\begin{equation}
\left\{
\begin{array}
[c]{cc}%
(u_{00}-u_{01})(1-\overline{m}_{t}\left(  z\right)  )+u_{01} & \text{if
}z<\overline{f}(t,r_{2}^{l})\\
g_{3}(\overline{m}_{t}\left(  z\right)  ;C_{5},C_{6}) & \text{if }\overline
{f}(t,r_{2}^{l})\leq z<\overline{f}(t,r_{1}^{l})\\
u_{2} & \text{if }\overline{f}(t,r_{1}^{l})\leq z<\underline{f}(t,r_{1}^{R})\\
g_{4}(\underline{m}_{t}\left(  z\right)  ;C_{7},C_{8}) & \text{if
}\underline{f}(t,r_{1}^{R})\leq z<\underline{f}(t,r_{2}^{R})\\
(u_{11}-u_{10})\underline{m}_{t}\left(  z\right)  +u_{10} & \text{if
}\underline{f}(t,r_{2}^{R})\leq z\text{.}%
\end{array}
\right. \label{v2}%
\end{equation}
Here $g_{3}$ and $g_{4}$ are identical to $g_{1}$ and $g_{2}$ (defined in
(\ref{g1g2})) respectively, except that the constants $C_{1},...,C_{4}$ are
replaced respectively by $C_{5},...,C_{8}$ given by%
\begin{align*}
C_{5}  & =-\hat{c}\ell(r_{1}^{l})\text{, \ }C_{7}=-\hat{c}\ell(r_{1}^{R})\\
C_{6}  & =u_{2}-\hat{c}[(2r_{1}^{l}-1)\log(\frac{r_{1}^{l}}{1-r_{1}^{l}}%
)-\ell(r_{1}^{l})r_{1}^{l}]\\
\text{\ }C_{8}  & =u_{2}-\hat{c}[(2r_{1}^{R}-1)\log(\frac{r_{1}^{R}}%
{1-r_{1}^{R}})-\ell(r_{1}^{R})r_{1}^{R}]\text{. }\hfill
\end{align*}

\noindent\textbf{Proof of (b)}: Since it is never optimal to choose $a_{2}$,
we can delete it from the set of feasible actions. The proof proceeds as in
(a.ii), though we define $v(t,z)=$%
\[
\left\{
\begin{array}
[c]{cc}%
(u_{00}-u_{01})(1-\overline{m}_{t}\left(  z\right)  )+u_{01} & \text{if
}z<\overline{f}(t,\tilde{r}^{l})\\
g_{5}(\overline{m}_{t}\left(  z\right)  ;C_{9},C_{10}) & \text{if }%
\overline{f}(t,\tilde{r}^{l})\leq z<\widetilde{z}_{t}\\
g_{6}(\underline{m}_{t}\left(  z\right)  ;C_{11},C_{12}) & \text{if
}\widetilde{z}_{t}\leq z<\underline{f}(t,\tilde{r}^{R})\\
(u_{11}-u_{10})\underline{m}_{t}\left(  z\right)  +u_{10} & \text{if
}\underline{f}(t,\tilde{r}^{R})\leq z,
\end{array}
\right.
\]
where $g_{5}$\ and $g_{6}$\ are identical to $g_{1}$\ and $g_{2}$\ (defined in
(\ref{g1g2})) respectively, except that the constants $C_{1},...,C_{4}$\ are
replaced respectively by $C_{9},...,C_{12}$\ given by
\begin{align*}
C_{9}  & =-\hat{c}\ell(\tilde{r}^{R})+u_{11}-u_{10}\\
C_{11}  & =-\hat{c}\ell(\tilde{r}^{l})+u_{01}-u_{00}\\
C_{10}  & =u_{10}-\hat{c}[1-\widetilde{l}(\tilde{r}^{R})]\\
C_{12}  & =u_{00}-\hat{c}[1-\widetilde{l}(\tilde{r}^{l})]\text{.}%
\end{align*}

\hfill

\subsection{Proof of Lemma \ref{lemma-r}\label{app-lemma-r}}

Define $\hat{l}(r)=(2r-1)\log(\frac{r}{1-r})$. We prove the existence and
uniqueness of solutions to the following equations:

\noindent\textbf{(\ref{equ1-critical point-2})}: Follows from
$l:(0,1)\rightarrow(-\infty,\infty)$ being surjective, continuous and strictly increasing.

\medskip

\noindent\textbf{(\ref{nonsymmetry-equ})}: Adapt the argument in Peskir and
Shiryaev (2006, p. 290) used for a classical optimal stopping problem,
generalized here to our context with ambiguity. For fixed $\hat{r}^{l}%
\in(0,\overline{\pi})$, consider the following equation for $V^{l}(y)$:%
\begin{equation}
\left\{
\begin{array}
[c]{rl}%
V^{l}(y) & =\hat{c}\hat{l}(y)+\hat{C}_{1}y+\hat{C}_{2}\\
V_{y}^{l}(y) & =\hat{c}l(y)+\hat{C}_{1}\\
V^{l}(\hat{r}^{l}) & =-(u_{00}-u_{01})\hat{r}^{l}+u_{00}\\
V_{y}^{l}(\hat{r}^{l}) & =u_{01}-u_{00}\text{,}%
\end{array}
\right. \label{eq-unique-l}%
\end{equation}
where $y\in(0,1)$ and $\hat{C}_{1}$, $\hat{C}_{2}$ are constants to be
determined. The solution is
\[
V^{l}(y)=\hat{c}\hat{l}(y)-(u_{00}-u_{01}+\hat{c}l(\hat{r}^{l}))y+u_{00}%
+\hat{c}(\hat{r}^{l}l(\hat{r}^{l})-\hat{l}(\hat{r}^{l}))\text{.}%
\]
Because $V^{l}(y)$ depends on $\hat{r}^{l}$, we denote the solution by
$V^{l}(y;\hat{r}^{l})$. If $V^{l}(\overline{\pi};\hat{r}^{l})<u_{00}$, then we
consider the following equation for $V^{R}(y)$:%
\begin{equation}
\left\{
\begin{array}
[c]{rl}%
V^{R}(y) & =\hat{c}\hat{l}(y)+\hat{C}_{3}y+\hat{C}_{4}\\
V_{y}^{R}(y) & =\hat{c}l(y)+\hat{C}_{3}\\
V^{R}(\underline{\pi}) & =V^{l}(\overline{\pi};\hat{r}^{l})\\
V_{y}^{R}(\underline{\pi}) & =V_{y}^{l}(\overline{\pi};\hat{r}^{l})\text{,}%
\end{array}
\right. \label{eq-unique-r}%
\end{equation}
where $y\in\lbrack\underline{\pi},1)$ and $\hat{C}_{3}$, $\hat{C}_{4}$ are
constants to be determined. The solution is%
\[
V^{R}(y)=\hat{c}\hat{l}(y)+(V_{y}^{l}(\overline{\pi};\hat{r}^{l})-\hat
{c}l(\underline{\pi}))y+V^{l}(\overline{\pi};\hat{r}^{l})+\hat{c}%
(\underline{\pi}l(\underline{\pi})-\hat{l}(\underline{\pi}))-\underline{\pi
}V_{y}^{l}(\overline{\pi};\hat{r}^{l})\text{.}%
\]
Denote the solution by $V^{R}(y;\hat{r}^{l})$. Since $\hat{l}^{\prime\prime
}(y)=l^{\prime}(y)>0$ for $y\in(0,1)$, it is easy to see that $V^{l}(y;\hat
{r}^{l})$ and $V^{R}(y;\hat{r}^{l})$ are strictly convex functions. Recall
that $\underline{\pi}=\underline{m}_{t}(\widetilde{z}_{t})$, $\overline{\pi
}=\overline{m}_{t}(\widetilde{z}_{t})$ and $\underline{\pi}(u_{11}%
-u_{10})+u_{10}=\left(  1-\overline{\pi}\right)  (u_{00}-u_{01})+u_{01} $.
Then, $V^{R}(\underline{\pi})=V^{l}(\overline{\pi};\hat{r}^{l})$ implies that
the function $y\longmapsto V^{R}(y;\hat{r}^{l})$ intersects $y\longmapsto
(u_{11}-u_{10})y+u_{10}$ for some $y\in(\underline{\pi},1)$ when $\hat{r}^{l}$
is close to $\overline{\pi}$. Let $y=\hat{y}^{l}$ satify $V^{l}(y;\hat{r}%
^{l})=u_{00}$. Then, $\hat{y}^{l}\downarrow0$ as $\hat{r}^{l}\downarrow0$.

Then, reducing $\hat{r}^{l}$ from $\overline{\pi}$ down to $0$ and applying
the properties established above, we obtain the existence of a unique point
$\hat{r}_{\ast}^{l}\in(0,\overline{\pi})$ for which there exists $\hat
{r}_{\ast}^{R}\in(\underline{\pi},1)$ such that
\begin{align}
V^{R}(\hat{r}_{\ast}^{R};\hat{r}_{\ast}^{l})  & =(u_{11}-u_{10})\hat{r}_{\ast
}^{R}+u_{10}\label{eq-unique-condition}\\
V_{y}^{R}(\hat{r}_{\ast}^{R};\hat{r}_{\ast}^{l})  & =u_{11}-u_{10}%
\text{.}\nonumber
\end{align}
Combining (\ref{eq-unique-l}), (\ref{eq-unique-r}) and
(\ref{eq-unique-condition}), we can verify that $(\hat{r}_{\ast}^{R},\hat
{r}_{\ast}^{l})$ is a solution of (\ref{nonsymmetry-equ}). Note that each step
of the derivation is reversible. Thus, there exists a unique solution
$(\widetilde{r}^{R},\widetilde{r}^{l})$ for (\ref{nonsymmetry-equ}).
Inequalities (\ref{rtilda}) follow directly from construction of the solution.

\medskip

\noindent\textbf{(\ref{equ1-critical point-1-L}) and
(\ref{equ1-critical point-1-R})}: By the definition of $u_{2}^{\ast\ast}$ and
equation (\ref{equ1-critical point-2}), it is easy to check that $u_{2}%
^{\ast\ast}>u_{01} $. Set $\hat{y}=\frac{u_{00}-u_{2}^{\ast\ast}}%
{u_{00}-u_{01}}$. Define the following payoff function%
\[
V(y)=\left\{
\begin{array}
[c]{lc}%
-(u_{00}-u_{01})y+u_{00}\text{ } & \text{if }y\in(0,\hat{y})\text{;}\\
u_{2}^{\ast\ast}\text{ } & \text{ if }y\in(\hat{y},1)\text{.}%
\end{array}
\right.
\]
Then arguing as in Peskir and Shiryaev (2006, p. 290), we can prove that there
exists a unique solution $(r_{2}^{l},r_{1}^{l})$ for
(\ref{equ1-critical point-1-L}). The proof for (\ref{equ1-critical point-1-R})
is similar. It is obvious that $r_{2}^{l}<r_{1}^{l}$ and $r_{1}^{R}<r_{2}^{R}$
due to $l$ being strictly increasing.

\medskip

Turn to the remainder of the lemma (we skip the most obvious assertions).
Given payoff symmetry, the definitions of $\overline{\pi}$ and $\underline{\pi
}$ imply that $\underline{\pi}+\overline{\pi}=1$. Then $r^{l}+r^{R}=1$ follows
from (\ref{equ1-critical point-2}) and $l\left(  r\right)  +l\left(
1-r\right)  =0$.

\medskip

\noindent\textbf{Prove (\ref{2equivalences})}: Verify that $\frac{1}%
{2}l\left(  r\right)  =\widetilde{l}\left(  r\right)  -\frac{1}{2r\left(
1-r\right)  }+1$ and rewrite (\ref{equ1-critical point-1-R}) as
\[%
\begin{array}
[c]{rl}%
\tilde{l}(r_{2}^{R})-\tilde{l}(r_{1}^{R}) & =\frac{1}{2r_{2}^{R}(1-r_{2}^{R}%
)}-\frac{1}{2r_{1}^{R}(1-r_{1}^{R})}+\frac{u_{11}-u_{10}}{\hat{c}}\\
\tilde{l}(r_{2}^{R})-\tilde{l}(r_{1}^{R}) & =\frac{u_{2}-u_{10}}{\hat{c}%
}\text{.}%
\end{array}
\]
{\Large \ }If $u_{2}=u_{2}^{\ast\ast}$, then, using payoff symmetry, we can
verify that $r_{2}^{R}=r^{R}$, $r_{1}^{R}=\underline{\pi}$ is the unique
solution of (\ref{equ1-critical point-1-R}). Next we prove that the solution
$r_{1}^{R}$ of (\ref{equ1-critical point-1-R}) is increasing with respect to
$u_{2}$. Note that $l^{\prime}(r)=\frac{1}{r^{2}(1-r)^{2}}$ and $\tilde
{l}^{\prime}(r)=\frac{1}{r(1-r)^{2}}$. From (\ref{equ1-critical point-1-R}),
derive%
\begin{align*}
l^{\prime}(r_{2}^{R})\frac{dr_{2}^{R}}{dr_{1}^{R}}-l^{\prime}(r_{1}^{R})  &
=0\\
\tilde{l}^{\prime}(r_{2}^{R})\frac{dr_{2}^{R}}{dr_{1}^{R}}\frac{dr_{1}^{R}%
}{du_{2}}-\tilde{l}(r_{1}^{R})\frac{dr_{1}^{R}}{du_{2}}  & =\frac{1}{\hat{c}%
}\text{.}%
\end{align*}
Thus,
\[
\frac{dr_{1}^{R}}{du_{2}}=\frac{(r_{1}^{R})^{2}(1-r_{1}^{R})^{2}}{\hat
{c}(r_{2}^{R}-r_{1}^{R})}>0\text{,}%
\]
which proves $r_{1}^{R}\geq\underline{\pi}\Longleftrightarrow~u_{2}\geq
u_{2}^{\ast\ast}$. Similarly, we can prove that $r_{1}^{l}\leq\overline{\pi
}~\Longleftrightarrow~u_{2}\geq u_{2}^{\ast\ast}$.\hfill$\blacksquare$%
\hfill\noindent

\subsection{Proofs for the applications\label{app-applications}}

\noindent\textbf{Proof of Theorem \ref{thm-ellsberg} }(Ellsberg): \noindent(i)
Compute that $\hat{c}=\frac{c\sigma^{2}}{2\alpha^{2}}$, $\widetilde{z}_{t}=0$,
$\underline{\pi}=\frac{1-\epsilon}{2}$, $\overline{\pi}=\frac{1+\epsilon}{2}$.
Equations (\ref{equ1-critical point-1-R}) and (\ref{equ1-critical point-1-L})
simplify to
\[%
\begin{array}
[c]{c}%
r_{2}^{R}+r_{1}^{R}=1,~l(r_{2}^{R})=\frac{2\alpha^{3}}{c\sigma^{2}}\\
r_{2}^{l}+r_{1}^{l}=1,~l(r_{1}^{l})=\frac{2\alpha^{3}}{c\sigma^{2}}\text{,}%
\end{array}
\]
(which exploit the fact that $u_{2}=\frac{1}{2}(u_{00}+u_{10})$), and the
functions $\overline{f}$ and $\underline{f}$ become
\begin{align*}
\overline{f}(t,r)  & =\frac{\sigma^{2}}{2\alpha}\log(\frac{1-\epsilon
}{1+\epsilon}\frac{r}{1-r})\\
\underline{f}(t,r)  & =\frac{\sigma^{2}}{2\alpha}\log(\frac{1+\epsilon
}{1-\epsilon}\frac{r}{1-r})\text{.}%
\end{align*}
If $r_{1}^{l}<\frac{1+\epsilon}{2}$, then $\overline{f}(t,r_{1}^{l})\leq
0\leq\underline{f}(t,r_{1}^{R})$. By Theorem \ref{thm-general}(a.i), the
signal $Z_{0}=0$ falls in the stopping region which leads to $\tau^{\ast}=0 $.
This proves (i) with $\widehat{r}=r_{1}^{l}$.

(ii) Equation (\ref{equ1-critical point-2}) becomes%
\[
r^{R}+r^{l}=1,~l(r^{R})+l(\frac{1+\epsilon}{2})=\tfrac{4\alpha^{3}}%
{c\sigma^{2}}\text{,}%
\]
and%
\[
\overline{z}\equiv\underline{f}(t,r^{R})=-\overline{f}(t,r^{l})=\frac
{\sigma^{2}}{2\alpha}\left[  \log(\frac{1+\epsilon}{1-\epsilon})+\log
(\frac{r^{R}}{1-r^{R}})\right]  .
\]
By Theorem \ref{thm-general}(a.ii), $\tau^{\ast}=\min\{t\geq0:$ $\mid
Z_{t}\mid\geq\overline{z}\}$.

Let $\overline{\overline{z}}$ be given by%
\[
\overline{\overline{z}}=\frac{\sigma^{2}}{2\alpha}\log(\frac{1+\epsilon
}{1-\epsilon})<\overline{z}\text{. }%
\]
It follows from (\ref{Xt}) and (\ref{Mt}) that at any given $t$, not
necessarily an optimal stopping time, betting on the ambiguous urn is
preferred to betting on the risky urn iff $\mid Z_{t}\mid\geq\overline
{\overline{z}}$. Thus at $\tau^{\ast}>0$, $\mid Z_{\tau^{\ast}}\mid
=\overline{z}>\overline{\overline{z}}$, and betting on the ambiguous urn is
optimal on stopping. $\ $

Finally, we show that $\overline{z}$ is increasing in $\epsilon$:
$\ell^{\prime}\left(  r\right)  =\frac{1}{r^{2}\left(  1-r\right)  ^{2}%
}\Longrightarrow\frac{d\overline{z}}{d\epsilon}>0$ iff \newline$\frac{2r^{R}%
}{1-\epsilon}\ell^{\prime}\left(  r^{R}\right)  >\frac{1+\epsilon}{1-r^{R}%
}\frac{1}{2}\ell^{\prime}\left(  \frac{1+\epsilon}{2}\right)  $ iff
$\ \frac{1+\epsilon}{2}\cdot\frac{1-\epsilon}{2}>r^{R}\left(  1-r^{R}\right)
$. But $\frac{1}{2}<\frac{1+\epsilon}{2}<r_{1}^{l}<r^{R}$ $\Longrightarrow
$\newline$\frac{1+\epsilon}{2}\cdot\frac{1-\epsilon}{2}>r_{1}^{l}\left(
1-r_{1}^{l}\right)  >r^{R}\left(  1-r^{R}\right)  $. This completes proof of
(ii) with $\overline{r}=r^{R}$.\hfill\ $\ \ \blacksquare$ \hfill

\medskip

\noindent\textbf{Proof of Theorem \ref{thm-test}} (hypothesis test): Given
Theorem \ref{thm-general}(b), it remains only to prove (\ref{r-ineqB})
assuming that $a=b$. Payoff symmetry implies that (\ref{nonsymmetry-equ})
reduces to (\ref{equ1-critical point-2}). Using also Lemma \ref{lemma-r},
conclude that $\tilde{r}^{l}=1-\tilde{r}^{R}$ and that $\tilde{r}^{R}$ solves
$l(\widetilde{r}^{R})=l(\underline{\pi})+\frac{b}{\hat{c}}<\frac{b}{\hat{c}}$.
For Bayesians, $\underline{\pi}=\overline{\pi}=\frac{b}{a+b}$, and
(\ref{r-tildaB}) implies that $\tilde{r}_{B}^{l}=1-\tilde{r}_{B}^{R}$ and
$l(\tilde{r}_{B}^{R})=\frac{a+b}{2\hat{c}}=\frac{b}{\hat{c}}$. Hence
$\tilde{r}^{R}<\tilde{r}_{B}^{R}$. $\blacksquare$

\end{document}